\documentclass[12pt,preprint]{aastex}

\def\Sec{\hbox{${}^{\prime\prime}$\llap{.}}}

\def\sec{\hbox{${}^{\prime\prime}$}}

\def\lae{\mathrel{<\kern-1.0em\lower0.9ex\hbox{$\sim$}}}
\def\gae{\mathrel{>\kern-1.0em\lower0.9ex\hbox{$\sim$}}}

\shorttitle{Novae in M49}
\shortauthors{Ferrarese et al.}

\begin{document}


\title{Hubble Space Telescope Observations of Novae in M49}


\author{Laura Ferrarese, Patrick C\^ot\'e and Andr\'es Jord\'an\altaffilmark{1}}
\affil{Department of Physics and Astronomy, Rutgers University, New Brunswick, 
NJ 08854; lff@physics.rutgers.edu; pcote@physics.rutgers.edu; andresj@physics.rutgers.edu}


\altaffiltext{1}{Claudio Anguita Fellow}


\begin{abstract}
A search for novae in M49 (NGC~4472) has been undertaken with the {\it
Hubble Space Telescope}. A 55-day observing campaign in F555W (19
epochs) and F814W (five epochs) has led to the discovery of nine
novae. We find that M49 may be under-abundant in slow, faint novae
relative to the Milky Way and M31. Instead, the decline rates of the
M49 novae are remarkably similar to those of novae in the LMC. This
fact argues against a simple classification of novae in ``bulge" and
``disk"  sub-classes. We examine the Maximum-Magnitude versus Rate of
Decline (MMRD) relation for novae in M49, finding only marginal
agreement with the Galactic and M31 MMRD relations. A recalibration of
the Buscombe--de Vaucouleurs relation gives an absolute magnitude 15
days past maximum of $M_{V,{\rm 15}} = -6.36\pm0.19$, which is
substantially brighter than  previous calibrations based on Galactic
novae.  Monte Carlo simulations yield a global nova rate for M49 of
$\eta = 100^{+35}_{-30}$~year$^{-1}$ and a  luminosity-specific nova
rate in the range $\nu_K = 1.7-2.5$~~year$^{-1}$~10$^{-10}$$L_{\rm K
{\odot}}$.  These rates are far lower than those predicted by current
models of nova production in elliptical galaxies and may point to a
relative scarity of novae progenitors, or an increased recurrence
timescale, in early-type environments.
\end{abstract}


\keywords{galaxies: individual (M49, NGC~4472) --- novae, cataclysmic
variables --- galaxies: star clusters --- stars: distances}


\section{Introduction}

As close binaries in which material is accreted onto the surface of a
white dwarf, novae form the cataclysmic variable (cv) subclass of variable
stars. With amplitudes of 10--20 magnitudes, they reach maximum magnitudes
of $-6.5 \lesssim M_V \lesssim -10$ soon after the onset of thermonuclear
runaway burning. These high luminosities --- coupled with their occurrence
in galaxies of all morphological types --- suggests that novae have
the potential to be useful distance indicators, provided that some aspect
of their behavior near maximum light can be used as a standard candle.

While the earliest observations of extragalactic novae were reported by Ritchey 
(1917) and Shapley (1917), it was Hubble's exhaustive study of M31 that probably
constituted the first dedicated search for novae in an external galaxy (Hubble 1929).
The full potential of novae as distance indicators 
became apparent following the discovery by Zwicky (1936) that their
peak brightness correlates with their rate of decline,
in the sense that bright novae fade more rapidly
than their faint counterparts. Although the nature and calibration of this 
``Maximum Magnitude versus Rate of Decline'' (MMRD) relation has
been investigated on many subsequent occasions ($e.g.$, McLaughlin 1945; Arp 1956;
Cohen 1985; Capaccioli et~al. 1990; Livio 1992; Downes \& Duerbeck 2000),
these investigations have usually relied on observations for a handful
novae in the Galaxy or in M31, which,
by virtue of its proximity and high luminosity, has remained the
premier target for extragalactic novae surveys
($e.g.$, Arp 1956; Rosino 1973; Rosino et~al 1989; Shafter \&
Irby 2001). 

Despite some notable exceptions ($e.g.$, Graham 1979; Pritchet \& van
den Bergh 1985; 1987), studies of extragalactic novae have remained
largely serendipitous in nature ($e.g.$, Ferrarese et~al. 1996).
While a few heroic attempts to detect and study novae in Virgo and
Fornax ellipticals
using ground-based telescopes (Pritchet \& van den Bergh 1985; 1987;
Shafter, Ciardullo \& Pritchet 2000; Della Valle \& Gilmozzi 2002)
yielded light curves of varying quality for a few novae in a handful
of galaxies, conclusions regarding the
universality of the MMRD relation and its potential as a distance
indicator rely almost entirely on observations of novae in the Galaxy,
M31 and LMC ($e.g.$, Della Valle \& Livio 1995). Accurate data for
a sample of novae belonging to an elliptical galaxy with a well known distance
would be invaluable in this regard, particularly since novae provide
one of the few direct probes of compact binaries in such environments.

In this paper, we report the results of the first dedicated search for
extragalactic novae with the {\it Hubble Space Telescope (HST)}.  {\it
HST} is an ideal instrument for such a survey thanks to its high
spatial resolution, its ability to reach faint magnitudes in
relatively short exposures, and the opportunity to schedule
observations according to  a pre-defined, optimized 
sequence.  Our target, M49 (NGC~4472), is an obvious choice for
several reasons. Not only it is the first ranked member of the Virgo
Cluster, it is also the optically brightest galaxy in the local
supercluster. Moreover, a variety of distance estimates are available
for both Virgo, and for M49 itself. In fact, M49 offers the
opportunity to compare, and perhaps even calibrate, the nova MMRD
relation directly against other Population II distance indicators such
as surface brightness fluctuations and globular clusters.

\section{Observing Strategy}
\label{sec:strategy}

The {\it HST} observations of M49 began on April 09, 2001 using the Wide
Field and Planetary Camera 2 (WFPC2) on board the Hubble Space
Telescope ({\it HST}).  A total of 38 F555W ($\sim$ Johnson $V$) images,
divided among 19 epochs, were obtained within a 55 day period, with
the sequence ending on June 03, 2001. For epochs 14 to 18 of the
sequence, F814W ($\sim$ Johnson $I$) images were taken immediately
following the F555W observations.  The relevant parameters of the
observations are summarized in Table~\ref{tab:log}.  All images
were obtained at  the same telescope roll angle. To facilitate removal
of cosmic rays, exposures taken at each epoch shared the same
pointing; however, the telescope was dithered by 0\Sec6 between
different epochs, to allow for accurate removal of CCD anomalies
(e.g. hot or dead pixels).  To maximize the number of novae outbursts,
the center of M49 was placed close  to the intersection
of the four WFPC2 chips (Figure~\ref{fig:dposs}).

The temporal sequence and duration of the exposures were chosen with
two main considerations in mind, namely: the need to produce well-sampled 
novae light curves (particularly near maximum light) and the need to
follow their decline for at least two magnitudes thereafter. Because
novae are not periodic (at least not on short timescales), there is
always an obvious gain in lengthening the duration of monitoring
programs. In practice, the length of our program was imposed by the
pragmatic requirement of monitoring the field without a change in
position angle, which {\it HST} can do for a maximum of about two
months. This is longer than the decline rate for most novae  (Capaccioli et al.
1989; Capaccioli et al. 1990; Downes \& Duerbeck 2000), and 
therefore adequate for our purposes.

Prior to outburst, nova progenitors are obviously much too faint to be detected at
the distance of M49.  Following outburst, they reach maximum light
rather quickly (typically within a few days) and then decline over
much longer periods: from several days to a few months. A two day
interval between visits was deemed sufficient to sample the light curve
adequately both before and immediately after maximum. In the interest
of keeping the total time request for the program within reasonable
limits, M49 was originally scheduled to be monitored every two days
with the F555W filter for the first 14 epochs (26 days), and
every five days thereafter for the final five epochs (25 additional
days) with both the F555W and F814W filters. Although novae outbursts
occurring during the second half of the sequence might not be
optimally sampled, it was hoped that the time and magnitude at maximum, as
well as the rate of decline, could be recovered with the aid of the color
information (Ferrarese et al. 1996; see also \S 4.1) and by employing
the well-sampled light curves of novae discovered during the first part
of the program as templates.

The observing strategy discussed above was disturbed by a telescope
safing event which occurred from 04/28/2001 to 04/30/2001. Because of
this interruption, what was planned to be the eleventh epoch of our
sequence was skipped and rescheduled at the end of the program, leaving an
unfortunate four day gap during the finely sampled portion of the
observing sequence.

\section{Data Analysis}

\subsection{Data Reduction}

The WFPC2 consists of four separate 800$\times$800 CCD detectors:
three Wide Field Camera (WFC) chips, with a pixel size of 0.10 arcsec
and a field of view of 1.3$\times$1.3 arcmin per chip, and one high
resolution Planetary Camera (PC1), with a pixel size of 0.046 arcsec
and a field of view of 36$\times$36 arcsec. The gain and readout noise
are about 7 e$^-$/DN and 5 e$^-$ respectively. Further details about
the instrument can be found in the {\it HST} WFPC2 Instrument Handbook
(Biretta et al. 2001). All of the M49 observations were obtained with
the telescope guiding in fine lock, which achieves a nominal pointing
stability of about 3 milliarcseconds.

The reduction of the M49 frames followed the standard pipeline
maintained by the Space Telescope Science Institute (STScI), and
included correction of small A/D errors; subtraction of a bias level
for each chip; subtraction of a superbias frame; subtraction of a dark
frame; correction for shutter shading effects and division by a flat
field. For each filter, two back-to-back exposures were obtained to
aid in the removal of cosmic rays. After standard reduction,
but prior to performing the
photometric analysis, these exposures were combined, and cosmic rays
flagged and removed by comparing the difference in values between
pairs of corresponding pixels to a local sigma calculated from the
combined effects of Poisson statistics and local noise. The final,
reduced, cosmic ray cleaned set of images consists of 19 F555W and
five F814W frames. As a final step prior to photometric reduction, the
vignetted edges of each chip were blocked and the geometric distortion
of the WFPC2 optics was corrected using a pixel area map, following
Stetson et al. (1998).

\subsection{Photometric Analysis}

Photometric analysis of the data was performed  using a variant of
DoPHOT (Schechter et al. 1993), developed specifically to handle the
peculiarities of the {\it HST}/WFPC2 Point Spread Function (PSF) (Saha et
al. 1994). As is standard procedure (e.g. Saha et al. 1996), DoPHOT was
first run on deep F555W and F814W ``template" frames made by coadding
all 38 F555W and 10 F814W images, respectively. To prevent DoPHOT
from triggering on an unreasonable number of noise spikes, a minimum
S/N = 4 was required for object detection. The resulting master
star list was then used as a position reference when DoPHOT was run on
each individual frame. Using this approach, the completeness limit of
the resulting photometry reaches fainter magnitudes than it would if
DoPHOT were to be run directly on each individual epoch without the
aid of a master star list. Furthermore, the results are more robust
against residual cosmic rays which might affect individual frames.

The major challenge posed to the photometry comes from the
fact that the vast majority of the point-like objects in the WFPC2
field are globular clusters belonging to M49's rich system. At the
distance of Virgo, globular clusters are marginally resolved, so that
their PSF differs from the stellar PSF (and indeed varies with
the size of the cluster). Because DoPHOT constructs its PSF directly from the
brightest and most isolated objects in the image, 
in the case of M49 such a PSF might not adequately
represent truly unresolved objects, such as novae. If this were the case,
then the nova photometry could be compromised since the photometric
corrections needed to convert the fitted DoPHOT magnitudes to conventional
magnitudes (see below) would be appropriate for the globular
clusters only.

Fortunately, this appears not to be the case. For a few of the epochs, 
the fitted DoPHOT magnitudes (which are simply related
to the height of the fitted PSF) were compared to those measured from test
runs in which DoPHOT was forced to use a fixed PSF, constructed  from {\it HST}/WFPC2
observations of the globular clusters Pal 4 and NGC 2419 (Stetson et
al. 1998).  No significant offsets or scale errors were measured for either
filter, indicating that the photometry is not affected by the small PSF
broadening driven by the globular clusters.

DoPHOT fit magnitudes differ from `conventional' magnitudes by an
additive aperture correction, which can be calculated as the
difference between fit magnitudes and aperture magnitudes, provided
that the latter can be obtained  for a sufficient number of bright,
isolated stars in the field. Since such stars are not available for M49,
a more robust approach is to use aperture corrections
measured from unrelated F555W and F814W WFPC2 observations of a
densely populated field in the Leo~I dwarf galaxy (Saha et
al. 1996). These images contain a multitude of stars across the entire
field, allowing to quantify the positional dependence of the aperture
corrections due to the spatial variations of the PSF. Conventionally,
aperture magnitudes are measured within a $9\sec \times 9\sec$
aperture; although it does not contain the total light from a star,
this aperture is large enough that  PSF changes across the field,
changes in focus over time,  and small changes in jitter from one
exposure to another do not affect the fraction of the total light that
lies within this aperture. Finally, the $m_{9\times9}$ magnitudes 
obtained in this way were converted to the `ground system' magnitudes F555W and
F814W as defined in Holtzman et al. (1995) using the zero points
derived from observations of $\omega$ Cen (Hill et al. 1998). 
In practice, this transformation is applied only to the results from the
deep, template frames.  Once the objects from the individual epochs
are matched to their counterparts  on the template, the offset for the
individual epoch is evaluated from the  ensemble average of magnitude
differences (object by object) between that particular epoch and the
template frame. 

Following Holtzman et al. (1995), F555W and F814W magnitudes can be
converted to $V$ and $I$ by applying a correction which  depends on
the color of the object under consideration. While color information
is not always available for the novae, this correction is always
smaller than 0.02 mag in $V$ and 0.03 mag in $I$ over the color
interval $0.2 < (V-I) < 0.7$ mag, a range that is  typical for novae
(see, $e.g.$, van den Bergh \& Younger 1987).  Although systematic,
this error is insignificant compared to the other sources of errors
which enter our analysis, and will be neglected for the remainder of
this paper.

\subsection{Variable Star Search}

Variability was searched for in the F555W frames using two independent
methods. The first method applies a  simple $\chi^2$ test to the
DoPHOT photometry for each object. The second method is similar to the
image subtraction technique described by Alard \& Lupton (1998), and
makes no use at all of the DoPHOT photometry.

For any object detected by DoPHOT in at least two of the F555W frames,
a reduced  $\chi^2_r$ was calculated as

$$\chi^2_r = {1 \over (n-1)} \sum_i^n{(m_i-\bar{m})^2 \over
\sigma_i^2}, \eqno(1)$$

\noindent where $m_i$ and $\sigma_i$ are the magnitude and rms error
of a particular object as measured in the $i-$th epoch, $\bar{m}$ is
the magnitude of the same object averaged over all epochs, and $n$ is the
number of epochs in which the object is detected. An object was
flagged as variable if $\chi^2_r \ge 3$; this is effectively
equivalent to selecting objects whose magnitude fluctuates around the
mean by $1.7 \sigma$ on average.

Most of the objects flagged are, in fact, not intrinsically variable;
rather, the large $\chi^2$ is triggered by various anomalies in the
images (e.g. residual cosmic ray events).  A visual inspection of the
light curve for each putative variable was sufficient to
identify and reject these cases. The remaining candidates were
visually inspected by blinking all of the individual frames against
each other. The final list consists of nine  bona-fide novae, all in the
WFC chips (\S 4). Aperture photometry performed for each nova on all
epochs in which the nova was detected provided an additional
confirmation as to the reliability of the DoPHOT photometry.

The image subtraction technique can be considered as a ``visual"
application of the $\chi^2$ method described above. First, all epochs
are shifted to a common reference frame. A standard deviation frame is
then created as:

$$stdev(x,y) = \sqrt{\sum_i{[f_i(x,y)-\bar{f}(x,y)]^2} \over 18},
\eqno(2)$$

\noindent where $f_i(x,y)$ is the number of counts detected at pixel
$(x,y)$ in the $i-th$ epoch F555W frame, and $\bar{f}(x,y)$ is the
mean flux at the same pixel, averaged over all epochs. To avoid hot
pixels and/or single cosmic ray hits that inflate the standard
deviation, at each pixel, the epoch with the largest value of
$f_i(x,y)$ is excluded from the summation in equation 2.  All objects
detected at the 2 $\sigma$ level relative to the mean local background in
the standard deviation frame were then visually inspected by blinking
all of the original frames against each other. This procedure
recovered all of the nine novae flagged by the $\chi^2$ test, but did
not produce any additional detections.

The location of the nine novae found in M49 is shown 
superimposed to gray scale images of each chip in Figures 2-4, and 
on an isophotal contour of M49 in Figure 5.
Zoomed-in snapshots of the novae around maximum light are
shown in Figures 6 and 7. The coordinates of the novae are given in
Table~\ref{tab:pos}, while Table~\ref{tab:phot1} lists the magnitudes
at each epoch. Light curves are shown in Figures 8-10.
 
\section{The Novae Light Curves}

Can we be sure that the nine variable objects are indeed novae
belonging to M49?  We immediately rule out the possibility that they
are Galactic RR Lyrae or Cepheid variables on the basis of their
distinctive light curves: none of the light curves is periodic, the
amplitude of variations would be atypically large for either class of
variables, and the variation time scale is too long for RR-Lyrae
variables. Furthermore, RR-Lyrae variables or Cepheids reaching
apparent magnitudes of $V \gtrsim 22$ would need to have distances
$\gtrsim$ 250 kpc, placing them outside  the Galaxy. Nor could the
variables be lower main sequence flare stars in the Galactic halo
since variability in such stars occurs rapidly, on timescales of
minutes to hours (Hoffmeister, Richter \& Wenzel 1984). The
possibility that the variables are distant AGNs is likewise untenable:
optical fluctuations of 1--3 magnitudes in  QSOs and Seyfert galaxies
typically happen on timescales of several months to years, rather than
weeks (Peterson 2001). Furthermore, there is no evidence for a
nonstellar appearance in any of the sources.

Distant supernovae are also unlikely candidates.  Over the redshift
range $0.3 \lesssim z \lesssim 1.2$, Type Ia supernovae will have
magnitudes of $20.5 \lesssim V \lesssim 26.5$ at peak brightness
($e.g.$, Schmidt et~al. 1998; Nobili et~al. 2003), falling by $\sim$
0.7 magnitudes about 15 days later ($e.g.$, Hamuy
et~al. 1996). Although  such distant supernovae would resemble novae
in M49, the number expected in our survey is far too low.  The rate of
faint, Type Ia supernovae (which, at these magnitudes, outnumber Type
II supernovae by a wide margin) is $\sim$~160~deg$^{-2}$~year$^{-1}$
over the range $0.3 \lesssim z \lesssim 1.2$ (Pain et~al. 2002). Our
WFPC2 field covers an area of 5.66 square arcminutes and the survey
duration is 55 days, giving a dismal $\sim$ 0.04 distant supernovae
expected in our survey.

More likely, the variables belong to M49 itself, as indicated  by the
fact that their spatial distribution (Figures 2-5) is not uniform
across the field, but concentrates towards the center of the galaxy.  
Based on the light curves, variation amplitudes, and the
fact that M49 is an E2 galaxy, we can again eliminate the possibility
of  long period variables or Cepheids.  Microlensing events of the
sort described by Baltz \& Silk (2000;~2001) are also unlikely.
Scaling from the event rate observed in M87, we expect approximately
one event per month for M49 (E. Baltz; private communication) with no
stipulation on timescale. From Figure~2 of Baltz \& Silk (2000), we
expect events of duration $t_{\rm FWHM} \gtrsim$ 5 days to comprise
only a small fraction of the overall event rate. Moreover,
microlensing light curves are expected to be  symmetric around the
peak time, which is not the case for any of  the five objects for
which our observations bracket the inferred maximum. We conclude that
microlensing is not a significant source of contamination, leaving
novae in M49 as the only realistic explanation for the observed flares.

Figure~\ref{fig:cumlight} shows a comparison between the cumulative
distribution for the novae and the underlying galaxy light, measured
in the $V-$band. As is apparent from Figures 2-5, the novae appear to
be more centrally concentrated than the galaxy light: a KS test shows
an 8\% probability that they were drawn from the same
distribution. Artificial star tests (\S\ref{sec:novarate}) predict
that only novae in the innermost $\sim 5$ arcsec of the galaxy will
elude detection, and that beyond 17 arcsec even novae which have
already declined by two magnitude at the time of discovery will be
recovered. Missing novae in the innermost region would of course only
exacerbate the contrast between the novae distribution and the
underlying galaxy light. Pritchet \& van den Bergh (1987) also found
that the cumulative distribution of novae in M49 does not follow the
underlying galaxy light, although in their case the novae seem to be
under-represented in the outer ($r > 40$\sec) parts of the
galaxy. These somewhat puzzling results deserve further investigation
in future surveys.

\subsection{Determination of Light Curve Parameters}
\label{sec:vmaxtmax}

As will be discussed in detail in \S\ref{sec:distind}, the use of
novae as distance indicators requires an accurate knowledge of their
light curves, in particular, the maximum magnitude, $V_{\rm max}$, the
time of maximum light, $t_{\rm max}$, and the time taken to decline
two magnitudes from peak brightness, $t_2$. Here we discuss the
determination of these parameters for our sample of novae.

{\it Novae \#1,2,3,8:} Unfortunately, the light curves for all of
these novae were already declining from maximum brightness at the
outset of our program, making direct measurements of $V_{\rm max}$ and
$t_{\rm max}$ impossible.  Instead, we take $V_{\rm max} \le V(1)$ and
$t_{\rm max} \le t(1)$, where $V(1)$ and $t(1)$ refer to the magnitude
and time of the first epoch. The time taken for the novae to decline
in brightness by two magnitudes, $t_2$, and the corresponding decline
rates, $\nu_d \equiv 2/t_2$ can be bracketed more securely, since the
observed light curves are very nearly linear for the first few
epochs. Both are therefore estimated directly from weighted
least-squares fits to the $V$-band light curves using epochs 1--5
(nova \#1), 1--6 (nova \#2), 1--3 (nova \#3), and 1--10 (nova \#8).

{\it Nova \#4:} The nova is detected both before and after maximum.
Although the entire light curve is only sampled at 5-day intervals,
both $V$ and $I$ data are available post-maximum, allowing a precise
determination of $t_{\rm max}$. According to van den Bergh \& Younger
(1987), novae at maximum light have ($B-V$)$_0$ = 0.23$\pm$0.06,
corresponding to a spectral type between A7 and F0. From Zombeck
(1990), this translates to ($V-I$)$_0$ = 0.37$\pm$0.10, or
($V-I$)$_{\rm max}=0.40\pm0.10$ in the case of M49, adopting a total
(foreground plus internal) reddening to the galaxy $A(V)=0.07$ and
the relative extinction coefficients from Schlegel, Finkbeiner \&
Davis (1998).   Linear extrapolations  from $V(14)$, $V(15)$, $I(14)$
and $I(15)$ then give $t_{\rm max} = 2452039.04\pm1.11$. The implied
magnitudes at this time are $V_{\rm max} = 23.33\pm0.13$ and $I_{\rm
max} = 22.93\pm0.03$, where the main contribution to the error comes
from the uncertainty in the intrinsic color at maximum. This nova has
not yet faded by two magnitudes by the end of our observing sequence,
but a weighted linear least-squares fit to $V(14)$--$V(19)$ ---
further constrained to pass through $V_{\rm max}$ --- gives $t_2 =
52.00\pm4.21$~days.

{\it Nova \#5:} As in the preceding case, the nova is detected during
its rise to maximum, and $V$ and $I$ light curves are available during
its post-maximum evolution. Extrapolating to ($V-I$)$_{\rm
max}~=~0.40\pm0.10$ on the basis of $V(15)$, $V(16)$, $I(15)$ and
$I(16)$ gives $t_{\rm max} = 2452041.42\pm1.95$, $V_{\rm max} =
23.10\pm0.52$ and $I_{\rm max} = 22.70\pm0.42$. A least-squares fit to
$V_{\rm max}$, $V(14)$ and $V(15)$ yields $t_2 = 7.53\pm1.14$~days.

{\it Nova \#6:} Although this nova is observed during its rise to
maximum, only limited color information is available during its
post-maximum evolution. We assume that peak brightness occurred midway
between epochs 8 and 9, so that $t_{\rm max} =
2452023.46\pm0.50$. Extrapolations based on $V(9),$ $V(10)$ and
$V(11)$ then yield $V_{\rm max} =   23.33\pm0.28$. A weighted
least-squares fit to $V(9)$--$V(15)$ gives $t_2 = 19.30\pm0.86$~days.

{\it Nova \#7:} Two observations of this nova are available  prior to
maximum light: $V(9)$ and $V(10)$. In an unfortunate example of
Murphy's law, the period around maximum coincided with the telescope
safing event discussed in \S\ref{sec:strategy} (the light curve of
nova 9 also suffered from the safing). Extrapolating forward from
$V(9)$ and $V(10)$, and backwards from $V(11)$ and  $V(12)$, gives
$t_{\rm max} = 2452027.33\pm0.50$ and $V_{\rm max} =
22.72\pm0.36$. Based on this $V_{\rm max}$, and interpolating between
$V(13)$ and $V(14)$, we find  $t_2 = 11.68\pm0.83$~days.

{\it Nova \#9:} In this case, there is some ambiguity concerning the
precise time of maximum. Given the short rise time for most novae, we
think it most likely that maximum brightness occurred shortly after
epoch 11. The other possibility --- that maximum brightness occurred
before epoch 11 and after epoch 10, when the nova was not detected ---
cannot be excluded given the fact that novae light curves do not
always exhibit a monotonic decline after maximum (see, $e.g.$, nova
\#6). Working under the assumption that peak brightness occurred
midway between epochs 11 and 12, at  $t_{\rm max} =
2452031.45\pm0.50$, extrapolating from $V(12)$ and $V(13)$ gives
$V_{\rm max} = 23.74\pm0.22$.  This nova never fades by two magnitudes
from $V_{\rm max}$ so we take $t_2 = 37.16\pm2.86$~days based on a
linear least-squares fit to $V(12)$--$V(19)$, constrained to pass
through $V_{\rm max}$.

Light curve parameters for all nine novae are summarized in
Table~\ref{tab:decline}.  From left to right, the columns of this
table record the identification number, the time of maximum light, the
$V$-magnitude at maximum, the absolute $V$ magnitude for ($m-M$) =
31.06$\pm$0.06 and $A(V)_g = 0.07$ (see \S 6.1), the time taken for
the light curve to decline by two magnitudes, and the corresponding
decline rate.

\section{The Distribution of Novae Decline Rates}
\label{sec:distrate}

It has been suggested on a number of occasions ($e.g.$, Duerbeck 1990;
Della Valle \& Duerbeck 1993; Della Valle 2002) that there exist two
populations of novae: bright, rapidly declining ``disk" novae and
faint, slowly declining ``bulge" novae. Since M49 is an elliptical
galaxy consisting mainly of old and intermediate-age stars ($e.g.$,
Trager et~al. 2000; Cohen et~al. 2003), one would expect our sample of
novae to consist entirely of faint objects with long decline times.
The cumulative fraction of novae as a function of decline rate,
$\nu_d = 2/t_2$ is shown in Figure~\ref{fig:cum}, compared to the
cumulative rates (again measured as  $\nu_d = 2/t_2$) observed for M31
(Capaccioli et al. 1989), the LMC (Capaccioli et al. 1990) and the
Milky Way (Downes \& Duerbeck 2000). The M49 sample seems to behave
exactly the opposite as expected: $i.e.$, faint objects with long
decline rates are under-abundant.  This is unlikely due to an
observational bias since our photometry is essentially complete down
to $V = 25.5$ mag. Even the faintest novae are expected to reach $23.8
\lesssim V \lesssim 24.5$ mag at maximum (see \S\ref{sec:mmrd}) and
decline by one magnitude or more during the length of our observing
period: if present, they should have been detected. A
Kolmogorov-Smirnov test yields significance levels of 21\% and 49\%
for the null hypothesis that the M49 novae sample and those of the
Milky Way and M31, respectively, are drawn from the same
population. However, the corresponding significance level for the M49
and LMC samples is 99\%.

It is worth noting that some of the decline rates for LMC and Milky
Way novae, and most of those for novae in M31, are based on
photographic magnitudes.  Since novae decline more slowly in $B$ than
$V$, the decline rates for the LMC, Milky Way and M31 novae should be
increased somewhat relative to those in M49, which are measured in the
$V$ band. According to Van den Bergh and Younger (1987), $\log t_2(V)
= 0.953 (\pm0.013) \log t_2(B)$. Correcting the Milky Way and LMC
samples produces no difference as far as the KS test is concerned. For
the M31 sample, it is not always clear which novae have photographic
magnitudes and which have $V-$ band magnitudes. In the extreme case in
which the decline rates for all novae are corrected, the disagreement
between the M49 and M31 samples becomes less severe, but a KS test
still returns a significance level of only 84\% that the two samples
are drawn from the same population.

The LMC novae are usually regarded as prototypical examples of young,
disk novae. We conclude that our survey provides no clear support for
a simple classification of novae into a disk and bulge populations,
although a larger sample of novae in additional early type galaxies is
needed for a clear resolution of the issue.

\section{Novae as Distance Indicators}
\label{sec:distind}

The main goal of our program is to assess the usefulness of novae as
standard candles and, if possible, to provide a calibration of the
various relations which could establish novae as reliable distance
indicators ($e.g.$, Livio 1992). Because such relations are, at
present, calibrated exclusively with novae in two spiral galaxies
($i.e.$, M31 and the Galaxy), their applicability to early type
galaxies remains untested. M49 was chosen for this program not only
because of its brightness, but also because of the availability of
secure distance estimates. Here, we focus on two Population II
distance indicators --- surface brightness fluctuations (SBF) and the
globular clusters luminosity function (GCLF).

\subsection{Distance Calibration via SBF and Globular Clusters}
\label{sec:distance}

Tonry et~al. (2001) report an SBF distance modulus of  ($m-M$)$_0$ =
31.06$\pm$0.10 for M49\footnotemark.  Since the best available Maximum
Magnitude versus Rate of Decline (MMRD) relation for novae is based
almost entirely on observations of M31, it is also of interest to know
the distance to M31 on the same SBF scale, to facilitate the
comparison of novae in the two galaxies. According to Tonry
et~al. (2001),  the SBF distance of M31 is ($m-M$)$_0$ =
24.40$\pm$0.08, giving a difference in distance moduli between the two
galaxies of $\Delta$($m-M$)$_0$ = 6.66$\pm$0.13 mag.

\footnotetext{The SBF calibration adopted by Tonry et al. is based on
Cepheid distances to nearby galaxies which use the same {\it
HST}/WFPC2 photometric calibration adopted for the novae in this
paper. Although the photometric scale of the WFPC2 has since been
refined (Stetson 1998), the novae photometry and SBF distance are
mutually consistent.}

The GCLF has been claimed to be a first class distance indicator
($e.g.$, Jacoby et al. 1992) although concerns have also been raised
(Ferrarese et al. 2000). Our co-added images of M49 are the deepest
ever obtained for this galaxy; outside of the central 20\sec, our deep
images reach $V \simeq 25$ with better than 90\% completeness. The
GCLF, $\phi(V)$, can therefore be derived directly from our data,
providing an independent check of the adopted SBF distance.

To do so, the co-added, cosmic ray cleaned, deep F555W and F814W
images (\S 3.1) were background-subtracted via multi-resolution
wavelet filtering using the MR/1 package (Starck, Bijaoui \& Murtagh
1998). Object detection and photometry on the background subtracted
frames was performed with SExtractor (Bertin \& Arnouts 1996), using a
detection threshold of three connected pixels above $3.5 \sigma$.
Object catalogs for the separate F555W and F814W frames were then
matched with a matching radius of two pixels, and calibrated $V$ and
$I$ magnitudes were obtained following Holtzman et~al (1995, see also
\S 3.2). The matched catalog was then trimmed to exclude objects with
$V < 25$ or colors outside the range $0.7 \le (V-I) \le 1.45$.

Our final catalog consists of 389 globular cluster candidates. For
comparison, Larsen et~al. (2001) have recently carried out a study of
the M49 globular cluster system based on three partially overlapping
fields. Using the same magnitude and color selection criteria as
described above, we find their catalog to contain a total of 661
objects; the larger sample is a consequence of the greater areal
coverage of their survey. A comparison of the M49 GCLFs derived in
this paper and Larsen et al. (2001) is given in Figure~\ref{fig:gclf}
(note that our luminosity function has been scaled upwards by the
ratio of the sample sizes, 661/389 $\approx$ 1.7). There is  good
agreement between the two luminosity functions.   Following the usual
procedure of parameterizing $\phi(V)$ as a Gaussian, we find best-fit
values for the turnover and dispersion to be $V^{\rm TO} =
23.87\pm0.06$ and $\sigma = 1.43\pm0.09$. With the calibration of
Harris (2001), this turnover corresponds to a distance modulus of
($m-M$) = 31.13$\pm$0.09.  This best-fit Gaussian is shown by the
solid curve in Figure~\ref{fig:gclf}. If we instead choose to fix the
location of the turnover based upon the SBF distance modulus of Tonry
et~al. (2001) and the $M_V^{\rm TO}$ calibration of Harris (2001), we
find $V^{\rm TO} \equiv 23.80$ and $\sigma = 1.42\pm0.07$. The
corresponding luminosity function is shown as the dashed curve in
Figure~\ref{fig:gclf}. We conclude that the SBF and globular cluster
luminosity function methods yield highly consistent results for the
distance of M49.

In what follows, we adopt the SBF M49 distance modulus of ($m-M$) =
$31.06\pm0.10$ mag, corresponding to a distance of 16.3$\pm0.7$ Mpc.

\subsection{The Maximum Magnitude versus Rate of Decline Relation}
\label{sec:mmrd}

Although a relation between the magnitude reached by novae at maximum
and their rate of decline was first proposed in 1936 by Zwicky, a
theoretical and phenomenological description of this MMRD relation has
proven remarkably elusive. From an observational standpoint, the main
obstacle remains the small number of galaxies for which large samples
of novae with high-quality light curves have been collected. The list
includes only three objects: M31, with $\approx$ 55 novae (Capaccioli
et al. 1989), the Galaxy, with a few dozen objects (Downes \& Duerbeck
2000), and the LMC, with about 10 novae (Capaccioli et al. 1990). In
many cases, analysis of the light curves is further complicated by the
fact that observations are available in non-standard, photographic
bandpasses.

These concerns notwithstanding, it is known that the Galactic MMRD
relation can be fit reasonably well with a power law (Cohen
1985). Based on new distances derived from expansion parallaxes for a
sample of Galactic novae, Downes \& Duerbeck (2000) find:

$$M_{V,max} = (-11.32 \pm 0.44) + (2.55 \pm 0.32) {\rm log}(t_2)
\eqno(3)$$

\noindent where $t_2$ is the time it takes for the nova to decline by
two magnitudes from peak brightness, $M_{V,max}$. Both the M31 sample,
and the combined M31 and LMC samples, are better described in terms of
a ``stretched'' S-shaped curve, whose most recent calibration is given
as (Della Valle \& Livio 1995):

$$M_{V,max} = -7.92 - 0.81 {\rm arctan} {{1.32 - {\rm log}t_2} \over
  0.23} \eqno(4)$$

Fitting a similar function to the sample of Galactic novae  produces a
zero point consistent with the one found for M31, but a significantly
higher contrast between faint and bright novae (Downes \& Duerbeck
2000), although it should be pointed out that the quality of the fit
is not significantly improved over that provided by the single power
law given in equation 3:

$$M_{V,max} = -8.02 - 1.23 {\rm arctan} {{1.32 - {\rm log}t_2} \over
  0.23}, \eqno(5)$$

It is important to note that there is no theoretical explanation for a
relation of the kind shown in equations 4 and 5. In fact, Downes \&
Duerbeck (2000) argue that a better, and physically motivated,
characterization of the Galactic MMRD relation might be expressed in
terms of a broken power-law, in which novae are divided in two
distinct subgroups on the basis of the shape of the light curve and
the detailed physics of the outburst (Duerbeck 1981).

Figures~\ref{fig:mmrdm31} and ~\ref{fig:mmrdmw} compare the location
of the nine M49 novae in the MMRD plane (circled points) with those of
M31 and the Milky Way, respectively. Absolute, dereddened magnitudes
and decline rates for the Milky Way novae are listed by Downes \&
Duerbeck (2000, Table 5), and can be easily scaled for comparison with
the M49 sample by adopting $(m-M)_0 = 31.06\pm0.10$ mag and $A(V)_g =
0.07$ for M49 (from Tonry et~al.  2001, using the relative extinction
coefficients of Schlegel, Finkbeiner \& Davis 1998). Decline rates and
apparent, reddened, photographic maximum magnitudes for the M31 novae
are from Table IV of Capaccioli et~al. (1989). The photographic
magnitudes can be transformed to the $V-$band according to the
relation $V \approx m_{pg} - 0.06$, but scaling the M31 novae samples
to the distance and reddening of M49 proves to be a non-trivial
task. Although the distance to M31 is well known (e.g. Ferrarese et
al. 2000, we adopt here ($m-M$)$_0$ = 24.40$\pm$0.08 mag from Tonry et
al. 2001), the total extinction to the galaxy is not.  The RC3 reports
values of $A(B)_{i}=0.67$ mag for the $B$-band extinction internal to
M31, based on the galaxy inclination and Hubble type. The foreground
(Galactic) reddening is generally taken as $E(B-V) = 0.08$ mag,
corresponding to $A(B) = 0.35$ mag (see Bianchi et al. 1996 for a
comprehensive discussion). Given these estimates, the total $V$-band
extinction to M31 is therefore $A(V)\sim0.77$ mag, where we have
adopted $A(B)/A(V) = 1.321$ following Schlegel, Finkbeiner \& Davis
(1998).  This value can be compared to the one derived by Bianchi et
al. (1996) using {\it HST} UV observations of eight blue
supergiants. Their study points to a total reddening to M31 in the
range $E(B-V)=0.13 - 0.20$ mag, which translates to a $V-$band total
extinction in the range $A(V) = 0.43 - 0.66$.  For consistency with
the absorption adopted for other galaxies with measured novae rates
(to be discussed in \S\ref{sec:novarate}), in what follows we adopt
$A(V)\sim0.77$ mag for M31, with a 20\% uncertainty.

Filled and open squares in Figure~\ref{fig:mmrdm31} represent novae
judged by Capaccioli et~al.  to have light curves of good and fair
quality, respectively.  Absolute magnitudes are shown on the right
vertical axis. The dashed curves in Figures~\ref{fig:mmrdm31} and
~\ref{fig:mmrdmw} represent the best fit to the M31 sample (from
Capaccioli et al. 1989), and the MW sample (Downed \& Duerbeck, 2000;
equation 5) respectively. Residuals between these relations and the
M49 novae are shown in the lower panels. The solid line in
Figures~\ref{fig:mmrdmw} is the best power law fit to the Galactic
sample (equation 3).

The M49 sample is formally consistent with the Milky Way sample, while
the agreement with the M31 data is less secure. Restricting ourselves
to the five M49 novae with measured $V_{\rm max}$ and $\nu_d$, the
mean offsets between M49 and the best fitting relations to the M31 and
Milky Way samples shown by the dashed lines in
Figures~\ref{fig:mmrdm31} and ~\ref{fig:mmrdmw} are  $-0.50 \pm 0.15$
(random) $\pm$ 0.20 (systematic) mag and $-0.18 \pm 0.15$ (random)
$\pm$ 0.10 (systematic) mag respectively. In the above estimate, the
systematic error arises from the uncertainties in the distances and
reddenings to M31 and M49, while the random component arises from the
uncertainties in the magnitudes at maximum for the M49 novae. Although
the M49 and M31 samples disagree at the 2$\sigma$ level, larger
uncertainties in the internal extinction to the galaxy than considered
here are possible.  The root-mean square scatters of the M49 relation
around the M31 and MW MMRD relations are 0.74 mag and 0.85 mag
respectively; in both cases, there appears to be a systematic trend
for fast novae to be under-luminous, and slow novae to be
over-luminous compared to expectations, although the small number of
objects does not allow us to speculate any further.

A formal fit to the M49 sample, using the same relations given in
equations 4 and 5, allowing only the zero point to vary, and
accounting for errors in the magnitudes, produces:

$$M_{V,max} = (-8.27 \pm 0.59) - 0.81 {\rm arctan} {{1.32 - {\rm
log}t_2} \over 0.23}; {\rm ~rms}=0.82 {\rm ~mag} \eqno(6)$$

\noindent and

$$M_{V,max} = (-8.56 \pm 0.86) - 1.23 {\rm arctan} {{1.32 - {\rm
log}t_2} \over 0.23}; {\rm ~rms}=1.24 {\rm ~mag} \eqno(7)$$

\noindent respectively. The large scatter about these relations
suggests that, regardless of the details of the calibration, caution
is needed when deriving distances via the MMRD relation.

\subsection{The Buscombe and de Vaucouleurs Relation}

Buscombe \& de Vaucouleurs (1955) first noted that the absolute
magnitude 15 days after maximum light, $M_{15}$, appears to be roughly
the same for all novae. The existence of a time, or time interval,
after maximum  at which the absolute brightness of novae are
approximatively constant follows from the MMRD relation ($e.g.$, Shara
1981b) --- because brighter novae decline faster than fainter ones,
the light curves must intersect sometime after maximum.  The
calibration of $M_{15}$ as a standard candle has been attempted on
many occasions ($e.g.$, Shara 1981a; van den Bergh \& Younger 1987;
Cohen 1985; Downes \& Duerbeck 2000), yielding a rather wide range in
results.

Figure~\ref{fig:mconstant} shows light curves for the five novae in
our sample (\# 4, 5, 6, 7 and 9) for which both $V_{\rm max}$ and
$t_{\rm max}$ can be measured directly from their light curves (see
\S\ref{sec:vmaxtmax}).  Each light curve has been shifted in time so
that  maximum light occurs at $t \equiv 0$. Averaging the apparent
magnitude at $t = 15$ days gives $V_{15} = 24.77\pm0.19$,
corresponding to
$$M_{V,15} = -6.36\pm0.19~{\rm (random)} \pm0.10~{\rm
(systematic)}\eqno(8)$$ where the random uncertainty represent the
error in the mean, and the systematic uncertainty reflects the assumed
errors on M49's distance and extinction. The standard deviation about
this mean value is $\sigma = 0.43$.

This determination of $M_{V,15}$ is strongly at odds with the value
$M_{V,15} = -5.23\pm0.16$ found by van den Bergh \& Younger (1987)
from a sample of six Galactic novae with known expansion parallaxes.
It is barely consistent with the value of $M_{V,15} = -5.60\pm0.45$
found by Cohen (1985) from her study of 11 novae with spatially
resolved expansion shells, but it is in reasonable  agreement with the
value of $M_{V,15} = -6.05\pm0.44$ found by Downes \& Duerbeck (2000)
using the same method, but for a larger number of Galactic novae.

The lower panel of Figure~\ref{fig:mconstant} shows the mean novae
magnitudes and standard deviations as a function of time after
maximum. It is interesting to note that a rather small standard
deviation ($\sigma = 0.37$ mag) is found at maximum. The mean
magnitude at maximum for the five observed novae is $M_{V,max} =
-7.89\pm0.17$ (random) $\pm0.10$ (systematic), although we strongly
caution against  using this estimate as a distance indicator. For
instance, nova \#2 was $\approx$ 0.6 magnitudes brighter than this at
the onset of our program, at which time it was already declining from
maximum.

Is it reasonable to use $M_{V,15}$ as a distance indicator? We can
test this hypothesis on novae \#1, 2, 3 and 8, which were already
declining at the beginning of our program. If we assume that these
novae reach  $V = 24.77\pm0.19$ fifteen days after maximum, then using
the available datapoints (Table~\ref{tab:phot1}), we can linearly
extrapolate the light curve and infer $M_{V,max}$ and $t_{max}$. The
$V$ magnitudes at maximum found in this way are $20.87 \pm 0.46$ for
nova \# 1, $21.53 \pm 0.20$ for nova \# 2, $17.86 \pm 1.1$ for nova \#
3, and $23.08 \pm 0.19$ for nova \# 4\footnote{For this nova, the
outburst is predicted to have occurred $1.8\pm2.7$ days after the
first epoch.}. If these data are added to the maximum magnitude and
rate of decline of the five novae for which both parameters are
measured directly, and the MMRD relation given in equation (5) is fit
to the data, a distance modulus to M49 of $(m-M)_0 = 30.67 \pm 0.71$
is obtained (the error accounts only for the uncertainties in the M49
data, not in the MMRD calibration). This is certainly consistent with
Tonry et al. $(m-M)_0 = 31.06 \pm 0.10$, but hardly useful.

We conclude that the large standard deviation shown for $M_{V,15}$,
coupled with the rather poor agreement between the different
calibrations of $M_{V,15}$ based on samples of Galactic novae, suggest
that one should exercise considerable caution when estimating
distances with this method. This perhaps should come as no surprise
since, as noted by Jacoby et~al. (1992), several ``exceptional" novae
in M31 ($e.g.$, Arp 1, 2 and 3) offered clear counterexamples to the
universality of the Buscombe \& de Vaucouleurs relation.

\section{Nova Rates}
\label{sec:novarate}

The nova rate is a fundamental property of any stellar population and
represents a direct probe of the abundance of compact, mass-transfer
binaries in the host galaxy. There have been some recent claims that
the luminosity-specific nova rate varies along the Hubble sequence,
with late-type galaxies being more efficient producers of novae than
their early-type counterparts ($e.g.$, Della Valle et~al. 1994;
Yungelson, Livio \& Tutukov 1997; Della Valle 2002; c.f. Shafter
et~al. 2000). M49, as an E2 galaxy and the optically brightest member
of the local supercluster, presents an opportunity to measure the nova
rate in an extreme environment.

We use a Monte Carlo approach to measure the global nova rate in M49.
The difference in distance moduli between M31 and M49 is taken to be
${\Delta}$($m-M$) = 6.66, as discussed in \S\ref{sec:distance}. For an
assumed global nova rate, $\eta$, we randomly select novae having
maximum magnitudes and rates of decline given by the M31 relation of
Capaccioli et~al. (1989), displaced to the distance of M49. Light
curves are then simulated by calculating $V$-band apparent magnitudes
at the dates of the actual observations, assuming a linear decline
from maximum light.

For each simulated nova, we randomly assign a galactocentric  position
assuming that the local nova rate correlates linearly with stellar
mass density, and adopting the M49 surface brightness model presented
in C\^ot\'e et~al. (2003). Any novae falling outside the regions of
our survey are discarded. Artificial star experiments were carried out
to determine the level of photometric completeness as a function of
radius and magnitude. The results of these experiments were used to
determine, on a case-by-case basis, if the simulated nova would be
detected in our survey at the assigned radius and given its
instantaneous magnitude at each epoch. We consider a simulated nova to
be ``discovered" if it is detected at two or more epochs and shows a
variation in brightness that is larger than that expected on the basis
of its photometric errors. We carry out 1000 simulations at each
$\eta$, recording the number of times the number of detected novae in
the simulations matched the observed number of novae.  This process is
repeated for global nova rates varying between $\eta$ = 0 and 300
year$^{-1}$ in steps of five.

The results of this exercise are plotted in Figure~\ref{fig:novastat}.
The histogram shows the number of matches found in the Monte Carlo
simulations, plotted against global nova rate. The vertical lines
outline the best estimate of the global nova rate, $\eta =
100^{+35}_{-30}$~year$^{-1}$, where the quoted uncertainties refer to
68\% confidence limits determined directly from the simulations.

How does this nova rate compare to those found in the literature?  In
a $\sim 30$ day campaign, Pritchet \& van den Bergh (1987) visually
identified eight novae in M49. Not accounting for incompleteness or
detection biases, this yields a nova rate of $\eta = 160 \pm
57$~year$^{-1}$ (for $(m-M)_0 = 31.0$), consistent with the value
derived in this paper.  The only other (undisturbed) giant elliptical
galaxy for which a detailed search for novae has been undertaken is
M87, the second brightest member of the Virgo cluster.  On the basis
of multi-epoch, ground-based H$\alpha$ imaging, Shafter et~al. (2000)
measured a global nova rate of $\eta = 91\pm34$~year$^{-1}$ for this
galaxy.  Scaling our nova rate for M49 downwards by the ratio of the
$K$-band luminosities  of M49 and M87, we predict $\eta \simeq
70\pm23$~year$^{-1}$ for M87, fully consistent with the nova rate of
Shafter et~al. (2000). On the other hand, our predicted nova rate for
M87 is strongly at odds with the extreme rate of ``200 to 300"
year$^{-1}$ found for M87 by Shara \& Zurek (2002)\footnote{As cited
in Matteucci et~al. (2003).}.  For instance, if we turn the argument
around and scale the Shara \& Zurek (2002) nova rate for M87 to the
luminosity of M49, we would expect $280 \lesssim \eta \lesssim
425$~year$^{-1}$. Our Monte Carlo simulations reveal that such high
nova rates can be ruled out at better than 99.9\% confidence.

The complete sample of galaxies having measured nova rates encompasses
objects with widely different morphological types, luminosities and
star formation histories. Relevant information for these galaxies is
summarized in Table~\ref{tab:novarate}.  Columns (1-3) give the galaxy
name, morphological type from the NASA Extragalactic Database (NED)
and global nova rate. Sources for these nova rates are given in the
footnotes to the table.  Note that there is some disagreement over the
global nova rate in M33: Della Valle et al. (1994)  quote a value of
$\eta = 4.6\pm0.9$~year$^{-1}$, while Sharov (1993) argues for an {\it
upper limit} of $\eta \simeq 0.45$~year$^{-1}$.  

We follow the usual approach of normalizing the global nova rates by
$K$-band luminosity (which more closely traces the mass in evolved
stars). We consider two estimates for the total, dereddened $K$-band
magnitudes $M_{K,0}$. A recent 2MASS release (Jarrett et al. 2003)
lists total (extrapolated), reddened, $K$-band magnitudes for all
galaxies (Column 5 of Table~\ref{tab:novarate}), with the exception of
the LMC and SMC. We corrected the 2MASS magnitudes for both internal
and Galactic extinction, using the $B$-band absorption $A(B)_i$ and
$A(B)_g$ from the RC3 and Schlegel, Finkbeiner \& Davis (1998)
respectively (columns 7 and 8 of Table~\ref{tab:novarate}), and
$A(K)/A(B) = 0.085$ (Schlegel, Finkbeiner \& Davis 1998).  $M_{K,0}$
can also be calculated by correcting the total $B$-band magnitude
(from the RC3, column 4 of Table~\ref{tab:novarate}) for extinction,
and then applying a mean, dereddened $(B-K)_0$ color (column 6 of
Table~\ref{tab:novarate}, with references given in the footnotes). To
transform the dereddened $M_{K,0}$ to a luminosity in solar units, we
adopt the distance moduli given in column (9), and $M_{K,\odot} =
3.33$ mag (Cox 2000). Columns (10-11) give the luminosity-specific
nova rates, $\nu_K$, in the two cases in which the galaxy $K$-band
luminosity is calculated starting from the RC3 $B$-band magnitudes, or
adopted from the 2MASS analysis, respectively.

Before proceeding with an analysis of the behavior of $\nu_K$ among
the full sample of galaxies, we pause to consider the implications of
our measured nova rate for M49. Our best estimate of the
luminosity-specific nova rate in M49, $\nu_K =
1.71\pm0.61$~year$^{-1}$~10$^{-10}$$L_{\rm K {\odot}}$ ($\nu_K =
2.52\pm0.91$~year$^{-1}$~10$^{-10}$$L_{\rm K {\odot}}$ when using the
2MASS data), is far smaller than predicted by theoretical models of
nova production.  For instance, Matteucci et~al. (2003) have computed
galactic nova rates as a function of mass and star formation
history. Using the elliptical galaxy template models, and adopting a
luminous mass of ${\cal M} \simeq 8\times10^{11}{\cal M}_{\odot}$ for
M49 (C\^ot\'e et~al. 2003), we find the Matteucci et~al. (2003) models
to predict global nova rates in the range $780 \lesssim \eta \lesssim
900$~year$^{-1}$, nearly an order of magnitude larger than the
observed value of $\eta = 100^{+35}_{-30}$~year$^{-1}$.  Given the
size of the discrepancy --- and since the theoretical models were
calibrated to reproduce the global nova rate of $\eta \simeq
25$~year$^{-1}$ for the Milky Way --- reconciling the  models and
observations for M49 may require global differences in the properties
of the novae progenitors in the two galaxies. Obvious candidates would
include a reduced binary fraction in M49 relative to the Milky Way, or
longer recurrence timescales between outbursts.

Figure~\ref{fig:novarate} shows $\nu_K$ plotted as a function of
$K$-band luminosity for M49 (circled point) and each of the  galaxies
in Table~\ref{tab:novarate} (dots).  Note that M33 is plotted for both
the high nova rate of Della Valle et~al. (1994) and the upper limit of
Sharov (1993). Excluding this galaxy, the weighted mean nova rate is
$\langle \nu_K \rangle = 1.58\pm0.16$~year$^{-1}$~10$^{-10}$$L_{\rm K
{\odot}}$ when $M_{K,0}$ is derived from the RC3 $B$-band magnitudes,
and $\langle \nu_K \rangle =
2.41\pm0.27$~year$^{-1}$~10$^{-10}$$L_{\rm K {\odot}}$ when the 2MASS
data are used.\footnote{For comparison, the unweighted mean value are
$\langle \nu_K \rangle = 2.20\pm0.24$~year$^{-1}$~10$^{-10}$$L_{\rm K
{\odot}}$ and $\langle \nu_K \rangle =
2.66\pm0.30$~year$^{-1}$~10$^{-10}$$L_{\rm K {\odot}}$ respectively.}

The $\sim 35\%$ increase in the  luminosity-specific nova rates when
the 2MASS data are used is a consequence of the 0.2 mag systematic
difference between the two estimates of $M_{K,0}$, with the 2MASS
magnitudes being fainter. It is unclear which estimate of $M_{K,0}$ is
more reliable: compared to the 2MASS values, the RC3 $B$-band
magnitudes are more robust against small errors in the sky estimates,
but are very sensitive to extinction corrections, which can be very
uncertain in the case of spiral galaxies. In either case, the
luminosity specific nova rate in  M49 appears to be in good agreement
with the mean determined from the complete sample.  Indeed, with the
possible exception of the LMC, every galaxy for which a reliable
measurement of $\nu_K$ exists appears to be consistent with a
``universal" value of $\langle \nu_K\rangle \approx
1.6-2.4$~year$^{-1}$~10$^{-10}$$L_{\rm K {\odot}}$. These findings are
fully consistent with those of Shafter et al. (2000).

This constancy of $\nu_K$ has implications for theoretical models of
the galactic nova rates, and for possible dependences on morphological
type and star formation history. Calculations by Yungelson, Livio \&
Tutukov (1997) predict a $\sim$ 20-fold increase in the nova rate of a
stellar population which has formed stars continuously over a Hubble
time, compared to that of a stellar population which arose in  a
single burst 15 Gyr ago. While the precise star formation history of
M49 is unknown, we note that this galaxy is likely to have a
significant number of stars younger than 5--8 Gyr (Trager et~al.
2000; Cohen, Blakeslee \& C\^ot\'e 2003).  Thus, if we scale our
luminosity-specific nova rate for M49 to late-type systems like the
LMC and SMC, we would expect $\nu_K \sim
30$~year$^{-1}$~10$^{-10}$$L_{\rm K {\odot}}$ for these galaxies.
Microlensing surveys of these galaxies show that such extreme rates
can be ruled out with high confidence, despite the large uncertainties
in the measured values of $\eta$ and $\nu_K$.

\section{Conclusions}

We have presented the results of an {\it HST}/WFPC2 program designed
to  discover novae in M49.  Nine novae, five of which with fairly
complete ($i.e.$, covering both the pre- and post-maximum phases) and
well-sampled light curves, were discovered in a 55-day campaign. These
nine novae have been used to  examine the properties of novae in
early-type galaxies, measure the nova rate in M49, and assess the
potential of novae as distance indicators.  The main results of our
study are as follows:

\begin{itemize}

\item
Compared to the M31 and Galactic samples, M49 may be under-abundant in
slow, faint novae. Moreover, the distribution of novae decline rates
in M49 is statistically indistinguishable from that observed for
LMC. Bearing in mind the small sample of novae on which our discussion
is based, the M49 results seem to argue against a simple
classification of novae in a bright, fast, disk population (which
should be prevalent in the LMC) and a faint, slow, bulge population
(to which all of the M49 novae should belong).

\item
At a distance modulus of $31.06 \pm 0.10$ mag, measured both using SBF
and GCLF, the zero point of the Maximum Magnitude versus Rate of
Decline relation for the M49 novae is consistent with that derived
from a sample of two dozen Galactic novae, with distances determined
using expansion parallaxes. The agreement between the M49 and M31 MMRD
relations is less satisfactory, possibly owing to the large
uncertainty associated with M31 internal extinction (which affects the
maximum magnitude of the novae observed in this galaxy). In both
cases, there seems to be a substantial difference in the shape of the
MMRD relation in M49,  the Milky Way and M31.

\item 
The mean magnitude of the M49 novae 15 days after maximum is
marginally consistent only with one of three proposed calibrations
based on Galactic novae. Furthermore, the magnitudes of the M49 novae
seem to display a smaller scatter around maximum light than at 15 days
past maximum. Altogether, these results caution against an
indiscriminate use of novae as distance indicators.

\item
The global nova rate in M49 is  $\eta = 100^{+35}_{-30}$~year$^{-1}$,
corresponding to a luminosity-specific nova rate $\nu_K$ in the range
1.7--2.5 year$^{-1}$~10$^{-10}$$L_{\rm K {\odot}}$ (depending on the
adopted estimate for the $K$-band luminosity of the  galaxy). This
estimate accounts for observational incompleteness, due both to the
magnitude detection limits, and to the selection criteria adopted in
the detection of the variable stars. The value of $\nu_K$ measured for
M49 is inconsistent with the predictions of the theoretical models,
unless global differences are invoked between the novae progenitors in
M49 and the Milky Way (against which the models are calibrated). The
luminosity specific nova rate in M49 is fully consistent with that measured
in  all other galaxies for which data are available, with the possible
exception of the LMC.

\item
Last but not least, the valuable lesson learned from our program is
that, overall, obtaining reliable light curves for novae is not a
trivial task. Our program consumed 24 orbits of {\it HST} time and
lead to the discovery of nine novae. For comparison, 16 orbits of {\it
HST} time were used to discover 52 Cepheid variables and measure an
8\% distance to M100, also in Virgo (Ferrarese et al. 1996).  In
retrospect, a few changes to our observing strategy would have been
advisable. Color information would have been desirable at all epochs,
and a two-day interval between subsequent exposures over the entire
sequence would have aided in the measurement of the novae light  curve
parameters. For new programs, the Advanced Camera for Surveys (ACS)
would be more suitable than WFPC2 both because of the smaller pixel
size (reducing the background noise) and higher sensitivity.  Although
these changes would lead to a better characterization of the novae
light curves, they would entail a large program, likely requiring many
dozens of {\it HST} orbits per galaxy. Even then, the low
luminosity-specific nova rate, and the apparently large scatter in the
MMRD and Buscombe-de Vaucouleurs relations would ultimately limit the
usefulness of novae as distance indicators. SBF has been proven to be
a reliable --- and efficient --- indicator for early type galaxies,
while the GCLF has the potential of becoming one: they both seem more
worthwhile choices for measuring distances.

\end{itemize}

\acknowledgments

We thank Edward Baltz for useful discussions, and the referee, Allen
Shafter, for providing very useful comments.  Support for program
GO-8677 was provided through a grant from the Space Telescope Science
Institute, which is operated by the Association of  Universities for
Research in Astronomy, Inc., under NASA contract
NAS5-26555. P.C. acknowledges additional support provided by NASA LTSA
grant NAG5-11714. L.F. acknowledges additional support provided by NASA
through LTSA grant number NRA-98-03-LTSA-03. A.J. acknowledges
additional financial support provided by the National Science
Foundation through a grant from the Association of Universities for
Research in Astronomy, Inc., under NSF cooperative agreement
AST-9613615, and by Fundaci\'on Andes under project No.C-13442.

\clearpage

NOTE: FIGURES 1 TO 7 ARE NOT INCLUDED IN THIS SUBMISSION BECAUSE OF
SPACE LIMITATIONS. THEY ARE INCLUDED IN THE PDF VERSION OF THIS PAPER
WHICH CAN BE FOUND AT
http://www.physics.rutgers.edu/$\sim$lff/publications.html

\begin{figure}[p]
\caption{Digitized sky survey image of M49 showing the location and
orientation of our WFPC2 fields. The image measures
$6^{\prime}\times6^{\prime}$.
\label{fig:dposs}}
\end{figure}

\begin{figure}[p]
\caption{{\it HST}/WFPC2 image of M49. This F555W image shows the WF2
field, with the positions of novae \#1--5 indicated.
\label{fig:wf2}}
\end{figure}

\nopagebreak

\begin{figure}[h]
\caption{{\it HST}/WFPC2 image of M49. This F555W image shows the WF3
field, with the positions of novae \#6 and 7 indicated.
\label{fig:wf3}}
\end{figure}

\begin{figure}[p]
\caption{{\it HST}/WFPC2 image of M49. This F555W image shows the WF4
field, with the positions of novae \#8 and 9 indicated.
\label{fig:wf4}}
\end{figure}

\begin{figure}[p]
\caption{The location of the nine novae discovered in M49,
superimposed on an isophotal contour of the galaxy. The isophotes are
drawn at 0.5 mag/arcsec$^2$ intervals, from $\mu_V = 16$
mag/arcsec$^2$ to $\mu_V = 20.5$ mag/arcsec$^2$. The WFPC2 footprint
is also shown. The image is 2\Sec7 on the side, and has the same
orientation as Figure 1 (N at the top and E to the left).
\label{fig:isophotes}}
\end{figure}

\begin{figure}[p]
\caption{Appearance of novae \#1--5 at five different epochs. Each
panel shows a  $2\farcs7\times2\farcs7$ region centered on the nova,
in the F555W bandpass.  Epoch numbers are indicated in each panel.
\label{fig:postagestamps1}}
\end{figure}

\begin{figure}[p]
\caption{Appearance of novae \#6--9 at five different epochs. Each
panel shows a  $2\farcs7\times2\farcs7$ region centered on the nova,
in the F555W bandpass.  Epoch numbers are indicated in each panel.
\label{fig:postagestamps2}}
\end{figure}

\clearpage

\begin{figure}
\plotone{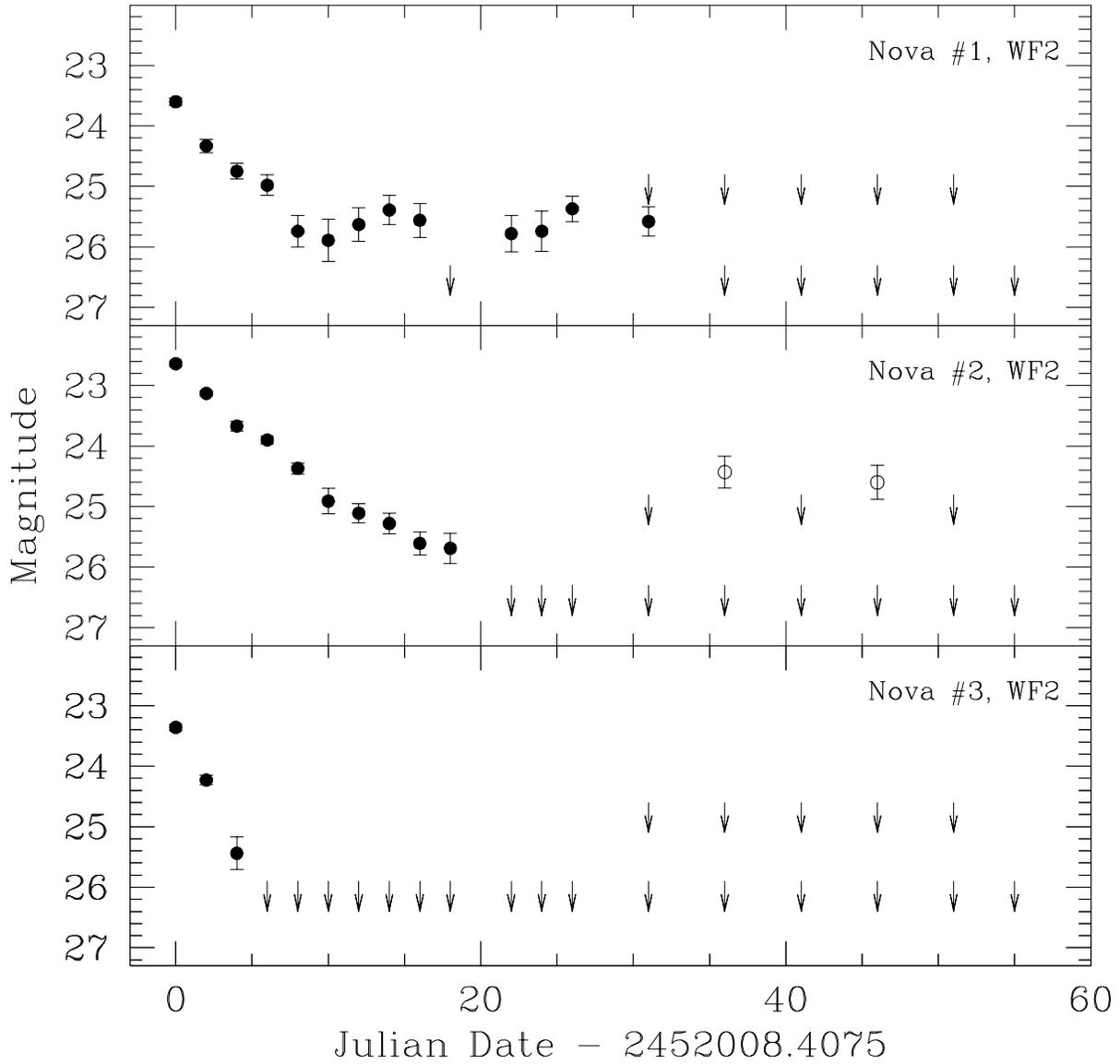}
\caption{Light curves for novae \#1, 2 and 3 in $V$ (filled symbols)
and $I$ (open symbols). The lower and upper arrows show respective
upper limits on the $V$ and $I$ magnitudes at epochs where the novae
were undetected.
\label{fig:lc1}}
\end{figure}

\clearpage

\begin{figure}
\plotone{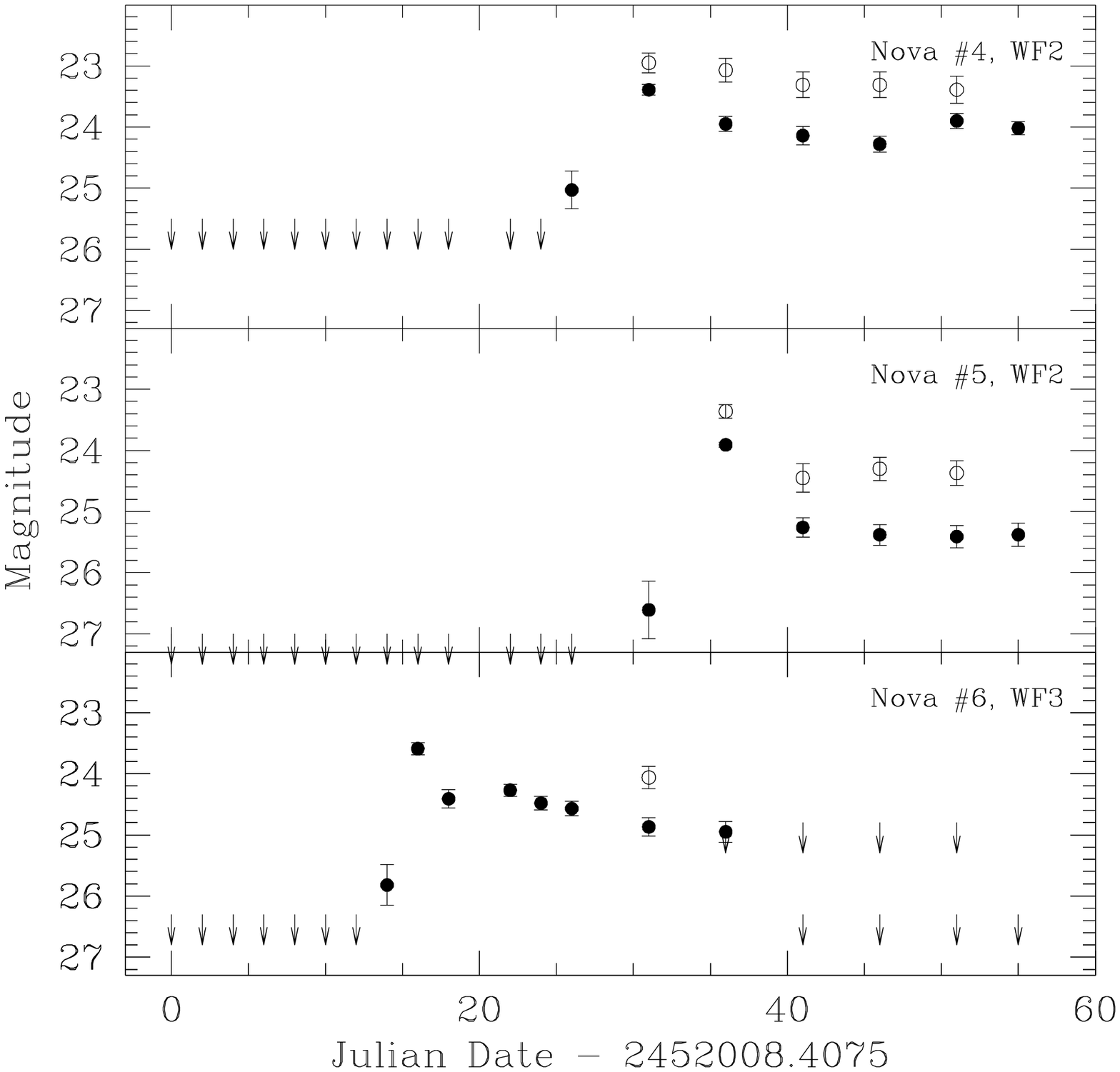}
\caption{Same as Figure~\ref{fig:lc1} except for novae \#4, 5 and 6.
\label{fig:lc2}}
\end{figure}

\clearpage
\begin{figure}
\plotone{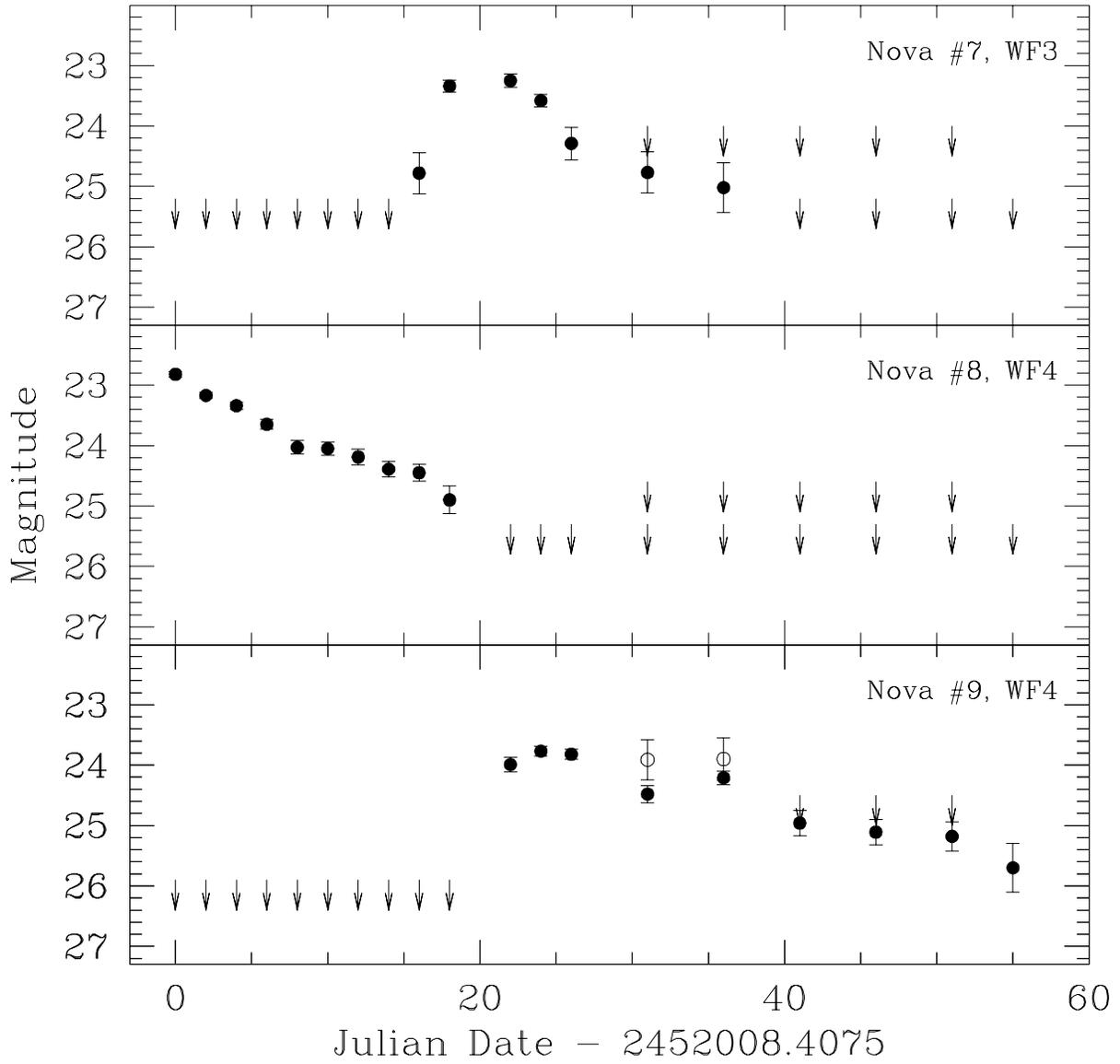}
\caption{Same as Figure~\ref{fig:lc1} except for novae \#7, 8 and 9.
\label{fig:lc3}}
\end{figure}

\clearpage

\begin{figure}
\plotone{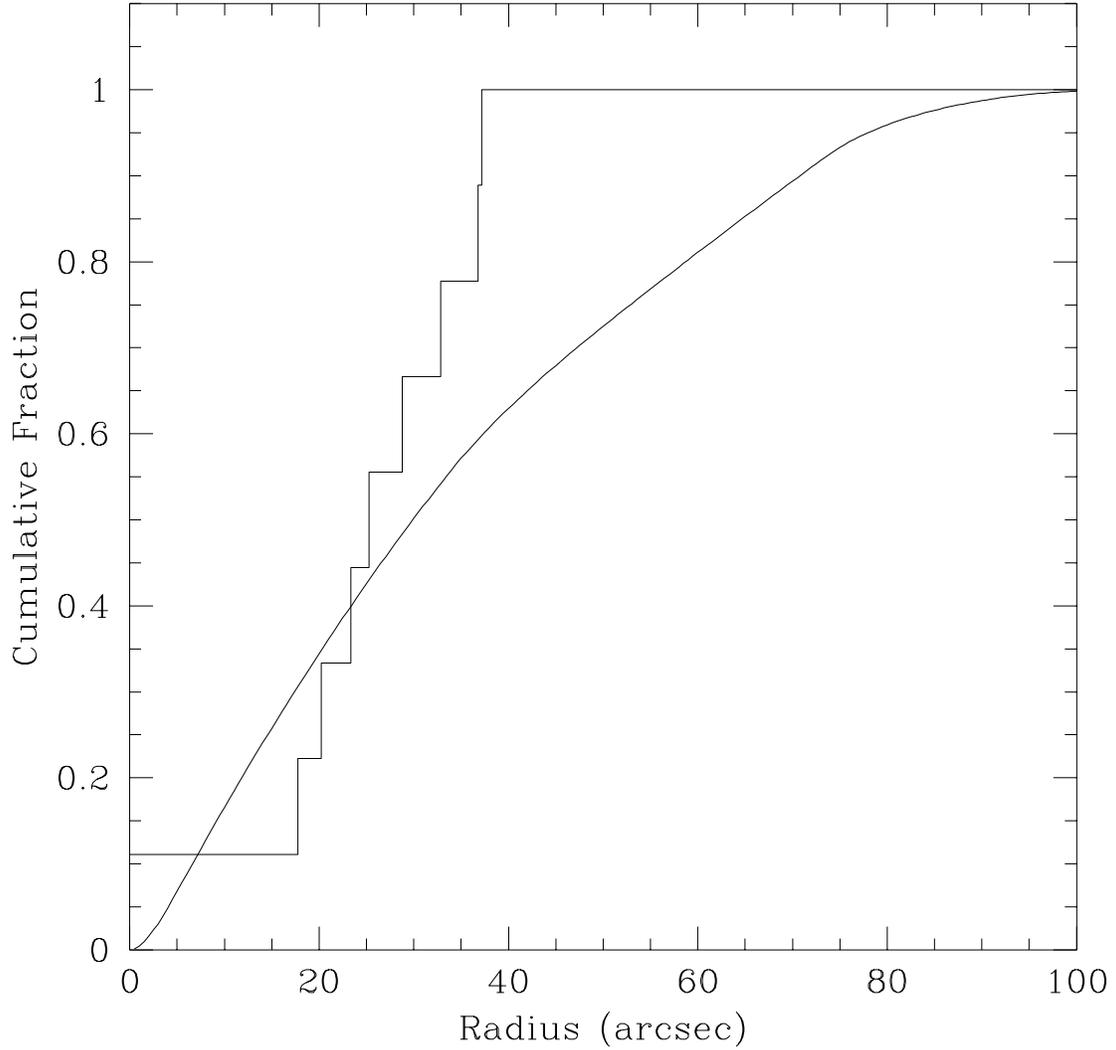}
\caption{A comparison between the cumulative distribution of the novae
detected in M49 and the cumulative fraction of the underlying galaxy
$V-$band light, normalized to the total light within the WFPC2
field. As is evident from Figures 2-5, the novae appear to be more
centrally concentrated than the galaxy light.
\label{fig:cumlight}}
\end{figure}

\clearpage

\begin{figure}
\plotone{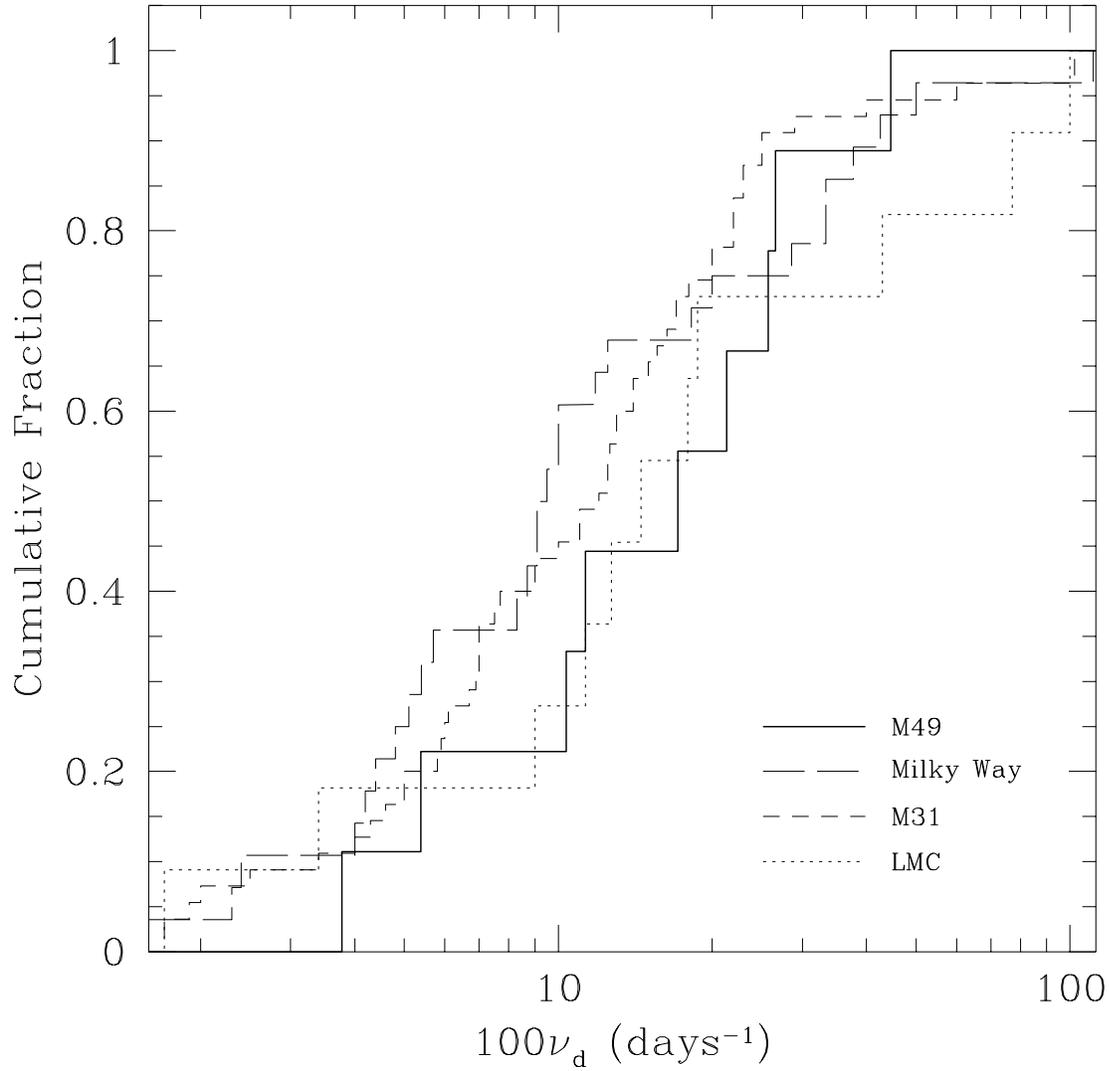}
\caption{The cumulative fraction of M31, LMC, Milky Way, and M49 novae
as a function of decline rate, defined as $\nu_d = 2/t_2$, where $t_2$
is the time it takes for the nova to decline by 2 magnitudes (in
$V$-band) after reaching maximum light. The LMC and M49 samples are
statistically indistiguishable.
\label{fig:cum}}
\end{figure}

\clearpage

\begin{figure}
\plotone{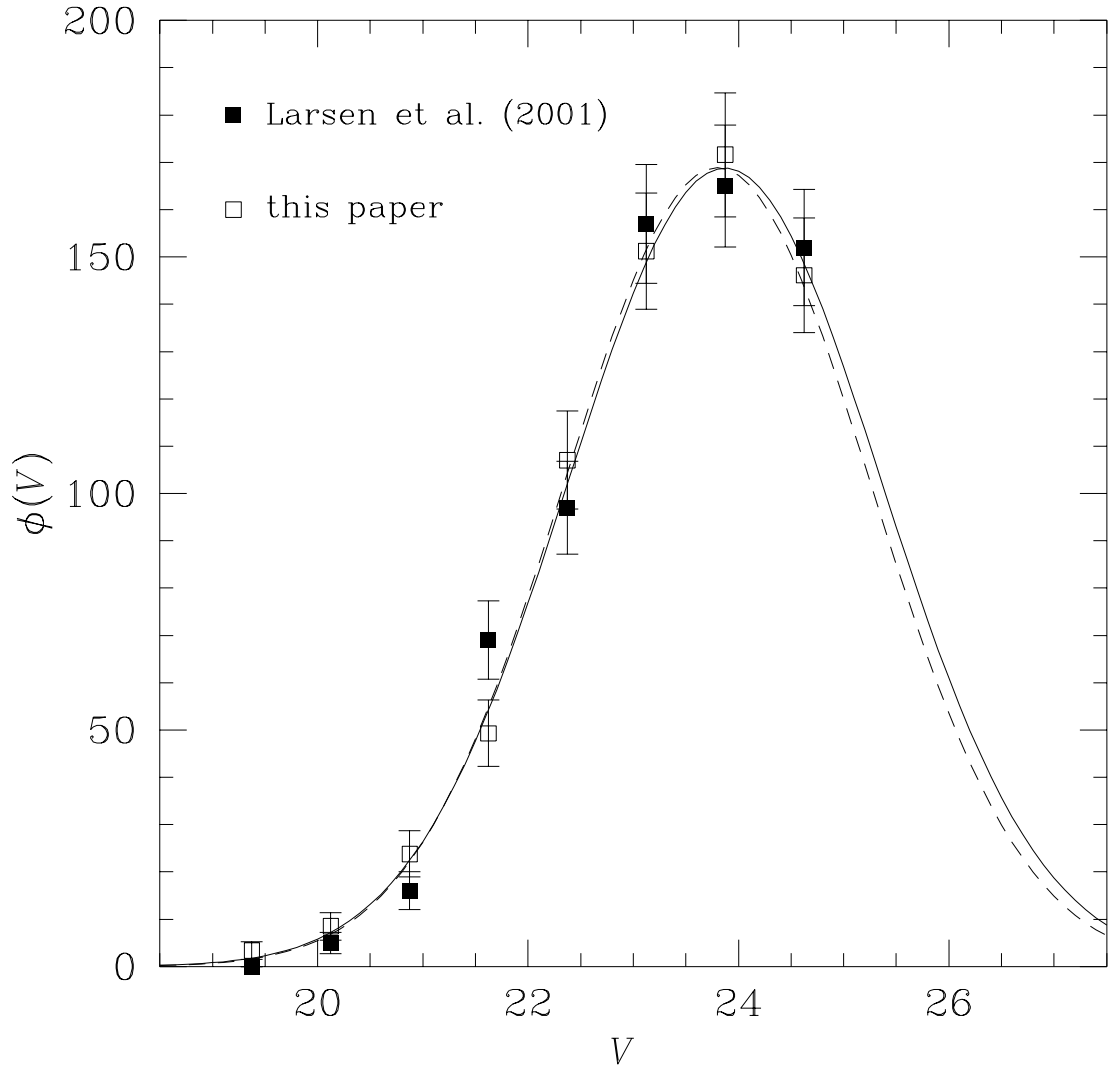}
\caption{ Globular cluster luminosity function, $\phi(V)$, for
M49. Filled symbols indicate the luminosity function of Larsen
et~al. (2001). Open symbols show the luminosity function determined
from our deep $V$ and $I$ images, scaled upwards by a factor (661/389)
$\approx$ 1.7. The solid curve shows an unconstrained Gaussian fit to
the luminosity function derived from our observations; the apparent
magnitude of the turnover is $V^{\rm TO} = 23.87\pm0.06$. The dashed
curve shows the best-fit obtained when the apparent turnover magnitude
is fixed at $V^{\rm TO} = 23.80$; this is the value expected on the
basis of the SBF distance modulus, ($m-M$)$_0$ = 31.06$\pm$0.10 (Tonry
et~al. 2001), and the absolute value of $M_V^{\rm TO} = -7.33\pm0.04$
given in Harris (2001).
\label{fig:gclf}}
\end{figure}

\clearpage

\begin{figure}
\plotone{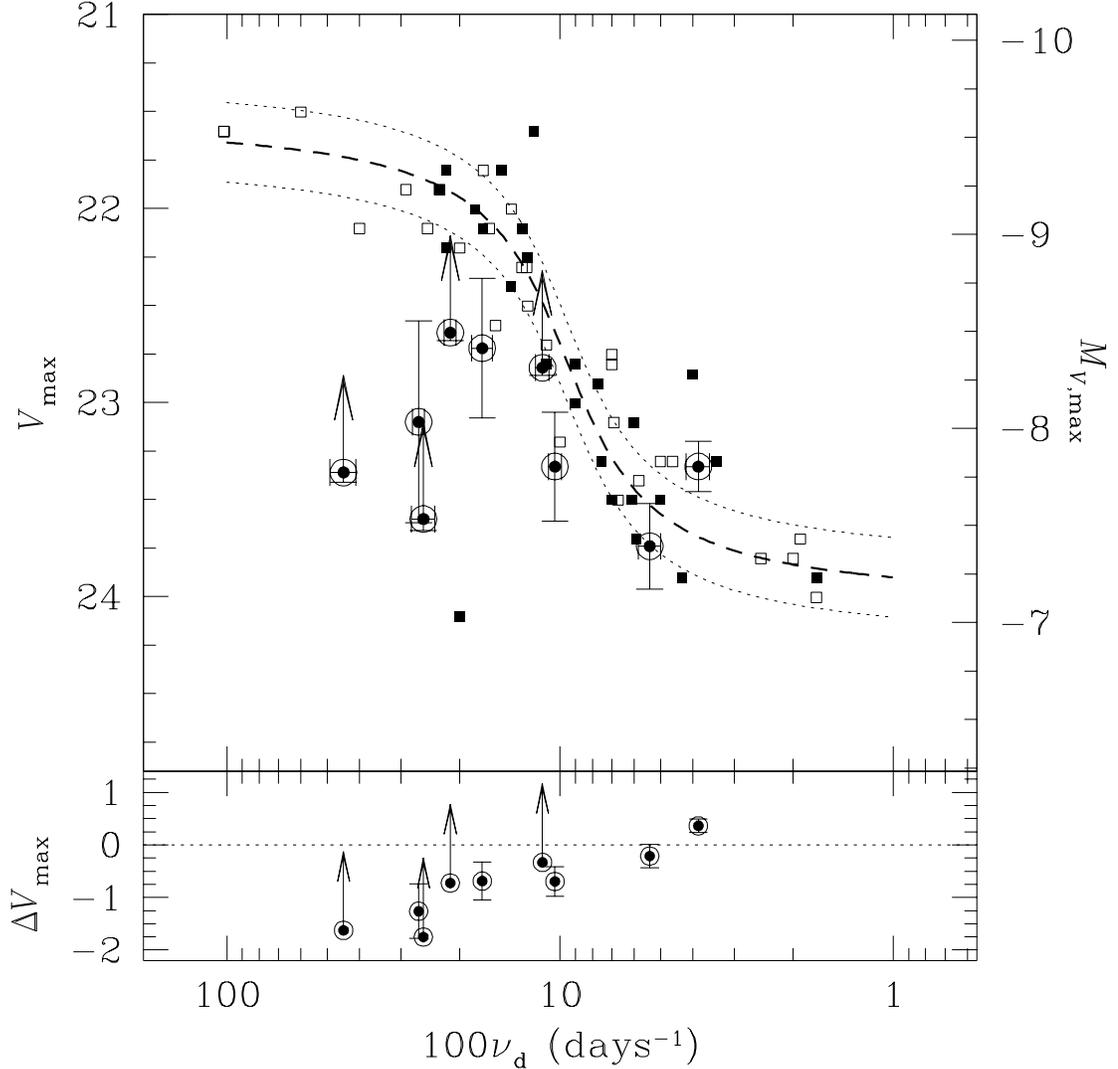}
\caption{The Maximum Magnitude versus Rate of Decline (MMRD) relation
for the M49 novae, shown as circled symbols. Filled and open squares
show M31 novae judged by Capaccioli et~al. (1989) to have light curves
of good and fair quality, respectively; their apparent magnitudes have
been scaled to the distance and reddening of M49. The thick dashed
curve shows the best fit MMRD relation for M31 (from Capaccioli et
al. 1989, again shifted to the distance of M49), with $1\sigma$
uncertainties shown by the thin dotted curves. Absolute magnitudes are
shown at right under the assumption that ($m-M$)$_0$ = 31.06$\pm$0.10
mag and $A(V)=0.07$ mag for M49. The lower panel shows the deviations
of the M49 novae from the M31 MMRD relation.
\label{fig:mmrdm31}}
\end{figure}

\clearpage

\begin{figure}
\plotone{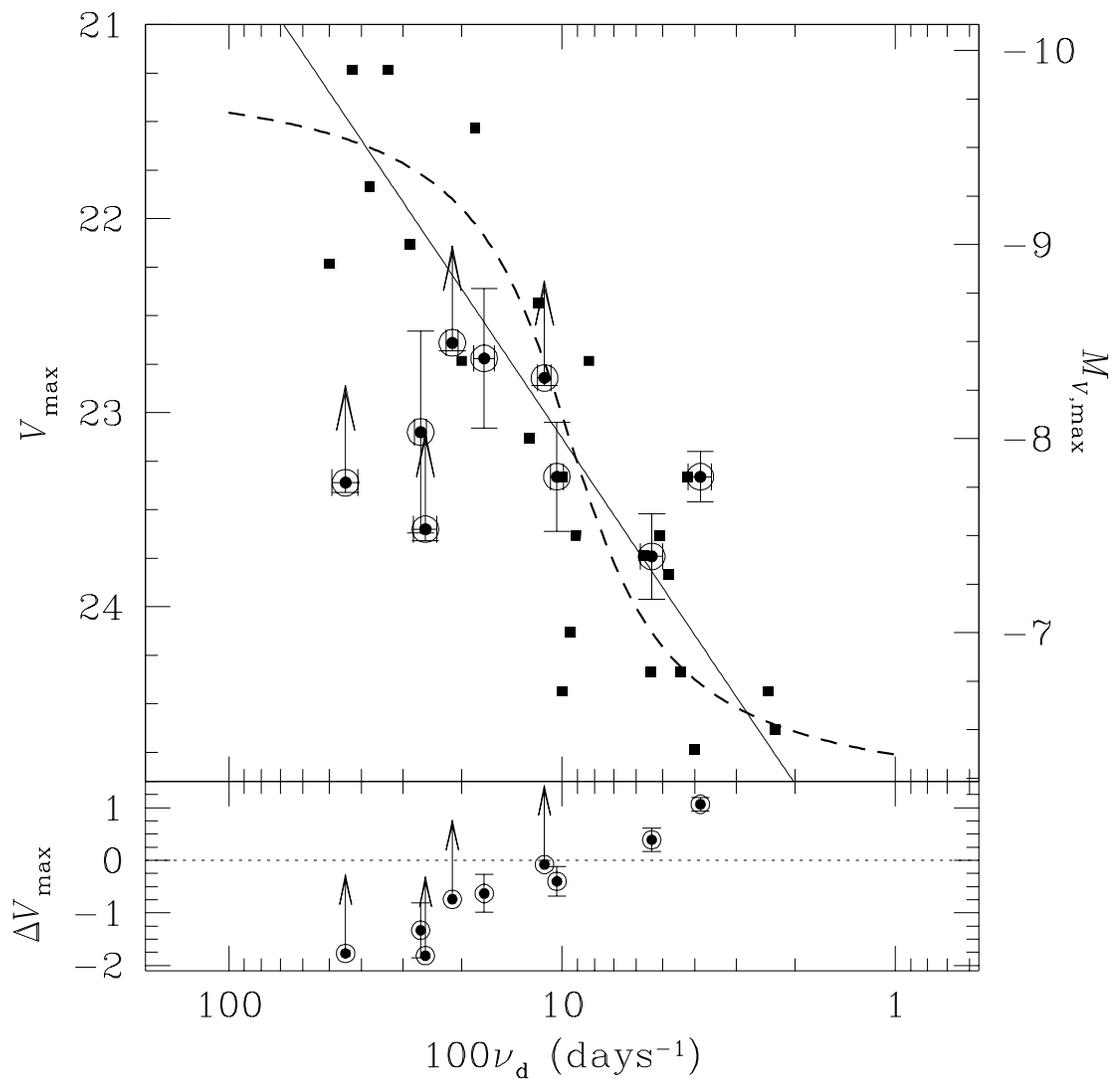}
\caption{Same as Figure~\ref{fig:mmrdm31}, except the comparison is
now between the M49 (circled symbols) and Galactic novae (filled
squares). The dashed and solid curves are the best fit ``S-shaped''
and power law fits to the Galactic novae from Downes \& Duerbeck
(2000); residuals between the former and the M49 novae are shown in
the lower panel.
\label{fig:mmrdmw}}
\end{figure}

\clearpage

\begin{figure}
\plotone{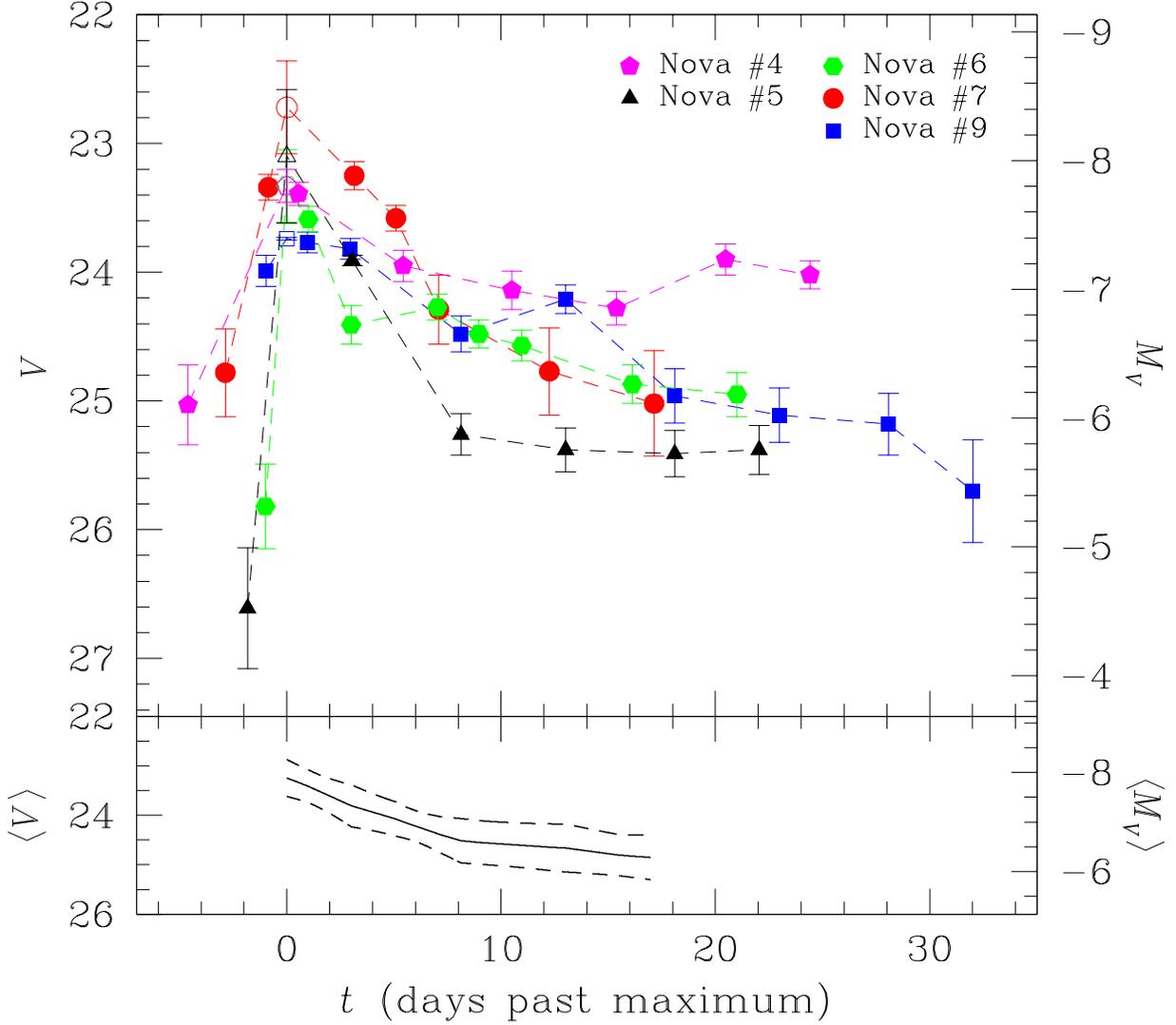}
\caption{{\it (Upper Panel)} Light curves for the five M49 novae
having measured $V_{\rm max}$ and $t_{\rm max}$. The light curves have
been shifted so that the times of maximum light align. Observed values
are shown as filled symbols; open symbols indicate our best estimates
for $V_{\rm max}$. The novae have $V_{15} = 24.77\pm0.19$ at 15 days
past maximum, corresponding to $M_{V,15}$ = $-$6.36$\pm$0.19. The
standard deviation about this mean value is $\sigma \simeq$ 0.43.
{\it (Lower Panel)} The mean magnitude averaged over the five novae
plotted in the upper panel, as a function of time following maximum.
\label{fig:mconstant}}
\end{figure}

\clearpage

\begin{figure}
\plotone{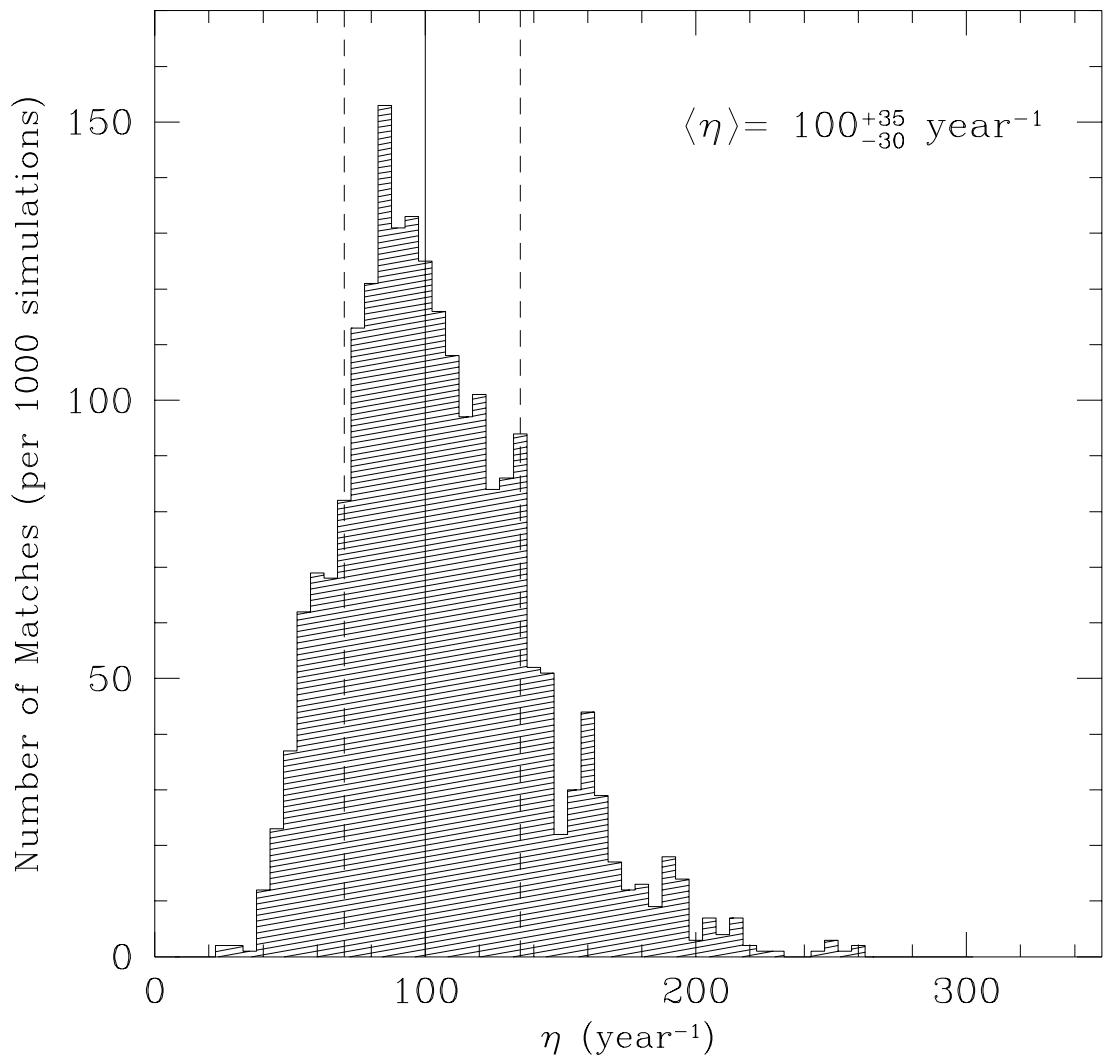}
\caption{Monte Carlo simulations of the global nova rate in M49. The
histogram gives the number of times --- based on 1000 simulations at
each $\eta$ --- that the simulated datasets contained exactly nine
novae that would have been discovered using our observing strategy and
data reduction procedures.  The solid and dashed lines show our best
estimates for the global nova rate and its 1-$\sigma$ uncertainties:
$\eta = 100^{+35}_{-30}$~year$^{-1}$.
\label{fig:novastat}}
\end{figure}

\clearpage

\begin{figure}
\plotone{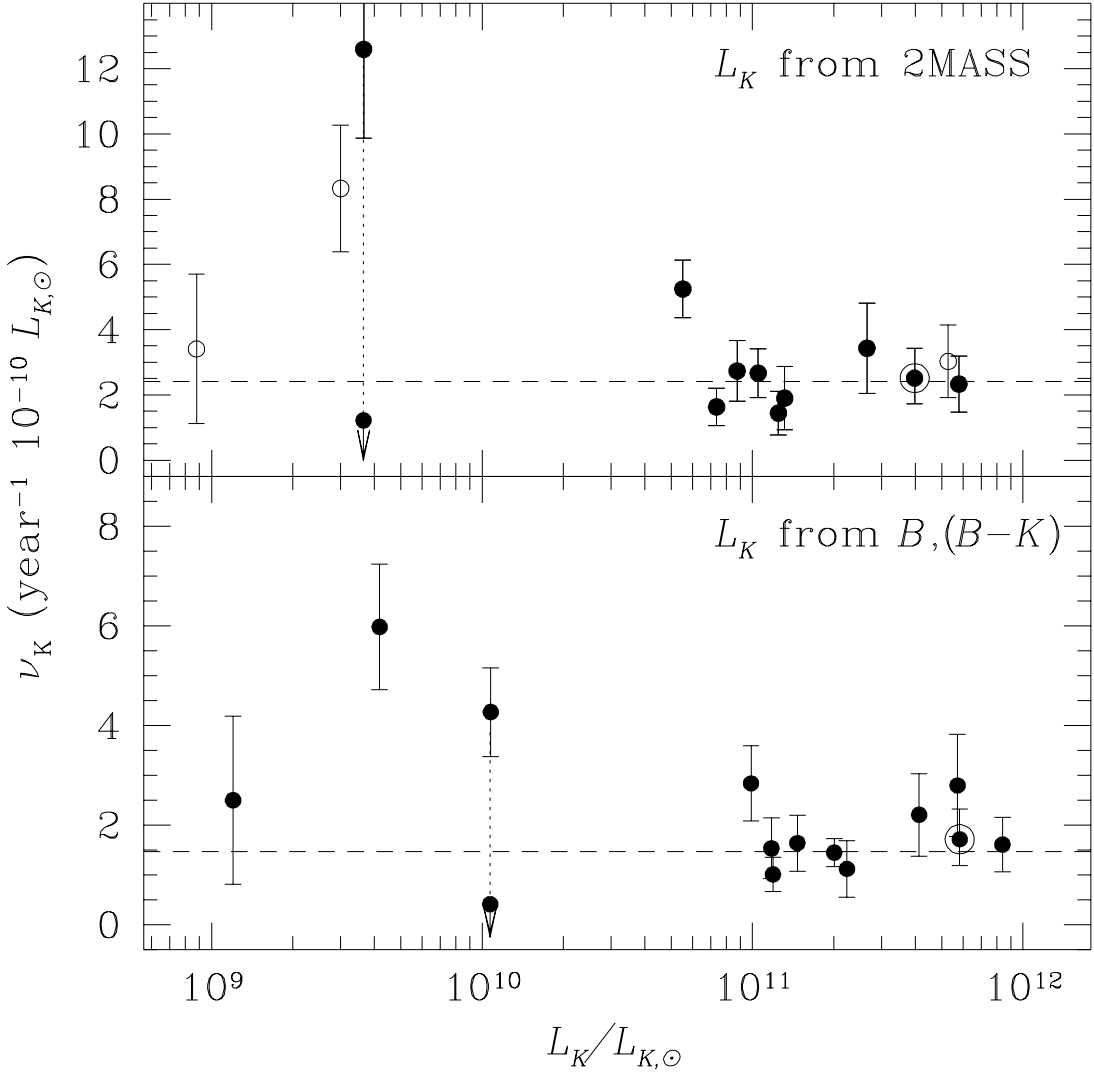}
\caption{Luminosity-specific nova rates, $\nu_K$, for all galaxies
with measured global nova rates.  The global nova rates have been
normalized by the total $K$-band luminosity of each galaxy, derived
from 2MASS (upper panel - the open circles denote galaxies for which
2MASS magnitudes are not available, and $K$-band magnitudes were
derived from the RC3 $B$-band magnitudes as in the lower panel) or
inferred from the RC3 $B$-band magnitudes via a color correction
(lower panel -- see text for further details). M49 is indicated by the
circled point. For M33, two different --- and inconsistent ---
estimates of the nova rate are available (see text for details). The
dashed line shows the weighted mean, exluding M33: $\bar{\nu_K} =
1.58\pm0.16$~year$^{-1}~10^{-10}L_{K,\odot}$ in the lower panel,
$\bar{\nu_K} = 2.41\pm0.27$~year$^{-1}~10^{-10}L_{K,\odot}$ in the
upper panel.
\label{fig:novarate}}
\end{figure}


\clearpage

\begin{deluxetable}{cccccc}
\tablecaption{Log of Observations GO-8677\label{tab:log}}
\tablewidth{0pt} \tablehead{ \colhead{Epoch} & \colhead{Observation
Date} & \colhead{Julian Date\tablenotemark{a}} & \colhead{Dataset
Name} & \colhead{Exposure Time}& \colhead{Filter} \\ \colhead{} &
\colhead{} & \colhead{(2,400,000+)} & \colhead{} & \colhead{(seconds)}
& \colhead{} } \startdata 1 & 2001 Apr  9 & 52008.39844 & U6630101R &
1200 & F555W \\ 1 & 2001 Apr  9 & 52008.41406 & U6630102R & 1200 &
F555W \\ 2 & 2001 Apr 11 & 52010.47266 & U6630201R & 1200 & F555W \\ 2
& 2001 Apr 11 & 52010.48438 & U6630202R & 1200 & F555W \\ 3 & 2001 Apr
13 & 52012.47656 & U6630301R & 1200 & F555W \\ 3 & 2001 Apr 13 &
52012.49219 & U6630302R & 1200 & F555W \\ 4 & 2001 Apr 15 &
52014.48438 & U6630401R & 1200 & F555W \\ 4 & 2001 Apr 15 &
52014.50000 & U6630402R & 1200 & F555W \\ 5 & 2001 Apr 17 &
52016.49219 & U6630501R & 1200 & F555W \\ 5 & 2001 Apr 17 &
52016.50781 & U6630502R & 1200 & F555W \\ 6 & 2001 Apr 19 &
52018.50000 & U6630601R & 1200 & F555W \\ 6 & 2001 Apr 19 &
52018.51562 & U6630602R & 1200 & F555W \\ 7 & 2001 Apr 21 &
52020.50391 & U6630701R & 1200 & F555W \\ 7 & 2001 Apr 21 &
52020.52344 & U6630702M & 1200 & F555W \\ 8 & 2001 Apr 23 &
52022.44141 & U6630801R & 1200 & F555W \\ 8 & 2001 Apr 23 &
52022.46094 & U6630802R & 1200 & F555W \\ 9 & 2001 Apr 25 &
52024.45312 & U6630901R & 1200 & F555W \\ 9 & 2001 Apr 25 &
52024.47266 & U6630902R & 1200 & F555W \\ 10 & 2001 Apr 27 &
52026.46094 & U6631001R & 1200 & F555W \\ 10 & 2001 Apr 27 &
52026.47656 & U6631002R & 1200 & F555W \\ 11 & 2001 May  1 &
52030.47656 & U6631201R & 1200 & F555W \\ 11 & 2001 May  1 &
52030.48828 & U6631202R & 1200 & F555W \\ 12 & 2001 May  3 &
52032.41016 & U6631301R & 1200 & F555W \\ 12 & 2001 May  3 &
52032.42578 & U6631302R & 1200 & F555W \\ 13 & 2001 May  5 &
52034.41797 & U6631401R & 1200 & F555W \\ 13 & 2001 May  5 &
52034.43359 & U6631402R & 1200 & F555W \\ 14 & 2001 May 10 &
52039.57422 & U6631501R & 1200 & F555W \\ 14 & 2001 May 10 &
52039.58594 & U6631502R & 1200 & F555W \\ 14 & 2001 May 10 &
52039.63281 & U6631503R & 1300 & F814W \\ 14 & 2001 May 10 &
52039.64844 & U6631504R & 1300 & F814W \\ 15 & 2001 May 15 &
52044.46094 & U6631601R & 1200 & F555W \\ 15 & 2001 May 15 &
52044.48047 & U6631602M & 1200 & F555W \\ 15 & 2001 May 15 &
52044.52734 & U6631603R & 1300 & F814W \\ 15 & 2001 May 15 &
52044.54297 & U6631604R & 1300 & F814W \\ 16 & 2001 May 20 &
52049.54297 & U6631701R & 1200 & F555W \\ 16 & 2001 May 20 &
52049.56250 & U6631702R & 1200 & F555W \\ 16 & 2001 May 20 &
52049.60938 & U6631703R & 1300 & F814W \\ 16 & 2001 May 20 &
52049.62500 & U6631704R & 1300 & F814W \\ 17 & 2001 May 25 &
52054.42578 & U6631801R & 1200 & F555W \\ 17 & 2001 May 25 &
52054.44531 & U6631802R & 1200 & F555W \\ 17 & 2001 May 25 &
52054.49219 & U6631803R & 1300 & F814W \\ 17 & 2001 May 25 &
52054.50781 & U6631804R & 1300 & F814W \\ 18 & 2001 May 30 &
52059.50781 & U6631901R & 1200 & F555W \\ 18 & 2001 May 30 &
52059.52344 & U6631902R & 1200 & F555W \\ 18 & 2001 May 30 &
52059.57812 & U6631903R & 1300 & F814W \\ 18 & 2001 May 30 &
52059.58984 & U6631904R & 1300 & F814W \\ 19 & 2001 Jun  3 &
52063.45312 & U6631101R & 1200 & F555W \\ 19 & 2001 Jun  3 &
52063.47266 & U6631102R & 1200 & F555W \\ \enddata
\tablenotetext{a}{Modified Julian date at the middle of the exposure.}
\end{deluxetable}

\clearpage

\begin{deluxetable}{cccc}
\tablecaption{Coordinates of Novae in M49\label{tab:pos}}
\tablewidth{0pt} \tablehead{ Nova   & $\alpha$(J2000) &
$\delta$(J2000) & $R$      \\ &                 &                 &
(arcsec) } \startdata 1 & 12:29:45.286 & 7:59:42.73 & 28.79 \\ 2 &
12:29:45.087 & 7:59:39.95 & 32.84 \\ 3 & 12:29:45.116 & 7:59:54.95 &
25.26 \\ 4 & 12:29:46.102 & 7:59:46.61 & 17.74 \\ 5 & 12:29:44.484 &
7:59:45.93 & 37.19 \\ 6 & 12:29:46.279 & 8:00:37.51 & 36.78 \\ 7 &
12:29:46.428 & 8:00:15.15 & 14.57 \\ 8 & 12:29:47.246 & 8:00:20.35 &
20.25 \\ 9 & 12:29:47.277 & 8:00:23.48 & 23.36 \\ \enddata
\tablecomments{Units of right ascension are hours, minutes, and
seconds,  and units of declination are degrees, arcminutes, and
arcseconds. $R$ is the distance of the nova from the center of M49.}
\end{deluxetable}

\clearpage

\begin{deluxetable}{cccccccccc}
\tablecaption{Novae Photometry\label{tab:phot1}} \scriptsize \rotate
\tablewidth{0pt} \tablehead{ Epoch & Julian Date &
\multicolumn{2}{c}{Nova 1} & \multicolumn{2}{c}{Nova 2} &
\multicolumn{2}{c}{Nova 3} & \multicolumn{2}{c}{Nova 4}\\ \cline{3-3}
\cline{4-5} \cline{6-6} \cline{7-8} \cline{9-10} & & $V$ & $I$ & $V$ &
$I$ & $V$ & $I$ & $V$ & $I$ } \startdata 1 &2452008.4075 &
23.60$\pm$0.06 & \nodata        & 22.64$\pm$0.04 & \nodata        &
23.36$\pm$0.05 & \nodata        & $\geq$25.50     & \nodata       \\ 2
&2452010.4818 & 24.33$\pm$0.11 & \nodata        & 23.13$\pm$0.04 &
\nodata        & 24.23$\pm$0.08 & \nodata        & $\geq$25.50    &
\nodata        \\ 3 &2452012.4887 & 24.75$\pm$0.13 & \nodata        &
23.67$\pm$0.08 & \nodata        & 25.44$\pm$0.27 & \nodata        &
$\geq$25.50    & \nodata        \\ 4 &2452014.4964 & 24.98$\pm$0.17 &
\nodata        & 23.90$\pm$0.06 & \nodata        & $\geq$25.90    &
\nodata        & $\geq$25.50    & \nodata        \\ 5 &2452016.5033 &
25.74$\pm$0.26 & \nodata        & 24.37$\pm$0.09 & \nodata        &
$\geq$25.90    & \nodata        & $\geq$25.50    & \nodata        \\ 6
&2452018.5102 & 25.89$\pm$0.35 & \nodata        & 24.91$\pm$0.21 &
\nodata        & $\geq$25.90    & \nodata        & $\geq$25.50    &
\nodata        \\ 7 &2452020.5179 & 25.63$\pm$0.28 & \nodata        &
25.11$\pm$0.16 & \nodata        & $\geq$25.90    & \nodata        &
$\geq$25.50    & \nodata        \\ 8 &2452022.4575 & 25.39$\pm$0.24 &
\nodata        & 25.28$\pm$0.17 & \nodata        & $\geq$25.90    &
\nodata        & $\geq$25.50    & \nodata        \\ 9 &2452024.4644 &
25.56$\pm$0.28 & \nodata        & 25.61$\pm$0.19 & \nodata        &
$\geq$25.90    & \nodata        & $\geq$25.50    & \nodata        \\
10 &2452026.4713 & $\geq$26.30    & \nodata        & 25.69$\pm$0.25 &
\nodata        & $\geq$25.90    & \nodata        & $\geq$25.50    &
\nodata        \\ 11 &2452030.4846 & 25.78$\pm$0.30 & \nodata        &
$\geq$26.30    & \nodata        & $\geq$25.90    & \nodata        &
$\geq$25.50    & \nodata        \\ 12 &2452032.4248 & 25.74$\pm$0.33 &
\nodata        & $\geq$26.30    & \nodata        & $\geq$25.90    &
\nodata        & $\geq$25.50    & \nodata        \\ 13 &2452034.4318 &
25.37$\pm$0.21 & \nodata        & $\geq$26.30    & \nodata        &
$\geq$25.90    & \nodata        & 25.03$\pm$0.31 & \nodata        \\
14 &2452039.5863 & 25.58$\pm$0.24 & $\geq$24.80    & $\geq$26.30    &
$\geq$24.80    & $\geq$25.90    & $\geq$24.60    & 23.39$\pm$0.09 &
22.95$\pm$0.16 \\ 15 &2452044.4706 & $\geq$26.30    & $\geq$24.80    &
$\geq$26.30    & 24.43$\pm$0.26 & $\geq$25.90    & $\geq$24.60    &
23.95$\pm$0.12 & 23.07$\pm$0.19 \\ 16 &2452049.5547 & $\geq$26.30    &
$\geq$24.80    & $\geq$26.30    & $\geq$24.80    & $\geq$25.90    &
$\geq$24.60    & 24.14$\pm$0.15 & 23.31$\pm$0.21 \\ 17 &2452054.4373 &
$\geq$26.30    & $\geq$24.80    & $\geq$26.30    & 24.60$\pm$0.28 &
$\geq$25.90    & $\geq$24.60    & 24.28$\pm$0.13 & 23.31$\pm$0.21 \\
18 &2452059.5207 & $\geq$26.30    & $\geq$24.80    & $\geq$26.30    &
$\geq$24.80    & $\geq$25.90    & $\geq$24.60    & 23.90$\pm$0.12 &
23.39$\pm$0.22 \\ 19 &2452063.4651 & $\geq$26.30    & \nodata        &
$\geq$26.30    & \nodata        & $\geq$25.90    & \nodata        &
24.02$\pm$0.11 & \nodata        \\ \enddata
\end{deluxetable}

\clearpage

\begin{deluxetable}{cccccccccc}
\tablecaption{Novae Photometry\label{tab:phot2}} \scriptsize \rotate
\tablewidth{0pt} \tablehead{ Epoch & Julian Date &
\multicolumn{2}{c}{Nova 5} & \multicolumn{2}{c}{Nova 6} &
\multicolumn{2}{c}{Nova 7} & \multicolumn{2}{c}{Nova 8} \\ \cline{3-4}
\cline{5-6} \cline{7-8} \cline{9-10} & & $V$ & $I$ & $V$ & $I$ & $V$ &
$I$ & $V$ & $I$ } \startdata 1 &2452008.4075& $\geq$27.00 & \nodata
& $\geq$26.30    & \nodata        & $\geq$25.20    & \nodata        &
22.82$\pm$0.04 & \nodata        \\ 2 &2452010.4818& $\geq$27.00    &
\nodata        & $\geq$26.30    & \nodata        & $\geq$25.20    &
\nodata        & 23.17$\pm$0.05 & \nodata        \\ 3 &2452012.4887&
$\geq$27.00    & \nodata        & $\geq$26.30    & \nodata        &
$\geq$25.20    & \nodata        & 23.34$\pm$0.06 & \nodata        \\ 4
&2452014.4964& $\geq$27.00    & \nodata        & $\geq$26.30    &
\nodata        & $\geq$25.20    & \nodata        & 23.65$\pm$0.08 &
\nodata        \\ 5 &2452016.5033& $\geq$27.00    & \nodata        &
$\geq$26.30    & \nodata        & $\geq$25.20    & \nodata        &
24.03$\pm$0.11 & \nodata        \\ 6 &2452018.5102& $\geq$27.00    &
\nodata        & $\geq$26.30    & \nodata        & $\geq$25.20    &
\nodata        & 24.05$\pm$0.11 & \nodata        \\ 7 &2452020.5179&
$\geq$27.00    & \nodata        & $\geq$26.30    & \nodata        &
$\geq$25.20    & \nodata        & 24.19$\pm$0.13 & \nodata        \\ 8
&2452022.4575& $\geq$27.00    & \nodata        & 25.82$\pm$0.83 &
\nodata        & $\geq$25.20    & \nodata        & 24.39$\pm$0.13 &
\nodata        \\ 9 &2452024.4644& $\geq$27.00    & \nodata        &
23.59$\pm$0.10 & \nodata        & 24.78$\pm$0.34 & \nodata        &
24.45$\pm$0.14 & \nodata        \\ 10 &2452026.4713& $\geq$27.00    &
\nodata        & 24.41$\pm$0.15 & \nodata        & 23.34$\pm$0.10 &
\nodata        & 24.90$\pm$0.23 & \nodata        \\ 11 &2452030.4846&
$\geq$27.00    & \nodata        & 24.27$\pm$0.10 & \nodata        &
23.25$\pm$0.11 & \nodata        & $\geq$25.30    & \nodata        \\
12 &2452032.4248& $\geq$27.00    & \nodata        & 24.48$\pm$0.11 &
\nodata        & 23.58$\pm$0.10 & \nodata        & $\geq$25.30    &
\nodata        \\ 13 &2452034.4318& $\geq$27.00    & \nodata        &
24.57$\pm$0.12 & \nodata        & 24.29$\pm$0.27 & \nodata        &
$\geq$25.30    & \nodata        \\ 14 &2452039.5863& 26.61$\pm$0.47 &
$\geq$25.10    & 24.87$\pm$0.15 & 24.06$\pm$0.18 & 24.77$\pm$0.34 &
$\geq$24.00    & $\geq$25.30    & $\geq$24.60    \\ 15 &2452044.4706&
23.91$\pm$0.04 & 23.36$\pm$0.11 & 24.95$\pm$0.17 & $\geq$24.80    &
25.02$\pm$0.41 & $\geq$24.00    & $\geq$25.30    & $\geq$24.60    \\
16 &2452049.5547& 25.26$\pm$0.16 & 24.45$\pm$0.23 & $\geq$26.30    &
$\geq$24.80    & $\geq$25.20    & $\geq$24.00    & $\geq$25.30    &
$\geq$24.60    \\ 17 &2452054.4373& 25.38$\pm$0.17 & 24.30$\pm$0.19 &
$\geq$26.30    & $\geq$24.80    & $\geq$25.20    & $\geq$24.00    &
$\geq$25.30    & $\geq$24.60    \\ 18 &2452059.5207& 25.41$\pm$0.18 &
24.37$\pm$0.20 & $\geq$26.30    & $\geq$24.80    & $\geq$25.20    &
$\geq$24.00    & $\geq$25.30    & $\geq$24.60    \\ 19 &2452063.4651&
25.38$\pm$0.19 & \nodata        & $\geq$26.30    & \nodata        &
$\geq$25.20    & \nodata        & $\geq$25.30    & \nodata        \\
\enddata
\end{deluxetable}
\clearpage

\begin{deluxetable}{cccc}
\tablecaption{Novae Photometry\label{tab:phot3}} \scriptsize
\tablewidth{0pt} \tablehead{ Epoch & Julian Date &
\multicolumn{2}{c}{Nova 9}\\  \cline{3-4} & & $V$ & $I$ } \startdata 1
&2452008.4075& $\geq$25.90    & \nodata        \\ 2 &2452010.4818&
$\geq$25.90    & \nodata        \\ 3 &2452012.4887& $\geq$25.90    &
\nodata        \\ 4 &2452014.4964& $\geq$25.90    & \nodata        \\
5 &2452016.5033& $\geq$25.90    & \nodata        \\ 6 &2452018.5102&
$\geq$25.90    & \nodata        \\ 7 &2452020.5179& $\geq$25.90    &
\nodata        \\ 8 &2452022.4575& $\geq$25.90    & \nodata        \\
9 &2452024.4644& $\geq$25.90    & \nodata        \\ 10 &2452026.4713&
$\geq$25.90    & \nodata        \\ 11 &2452030.4846& 23.99$\pm$0.12 &
\nodata        \\ 12 &2452032.4248& 23.77$\pm$0.08 & \nodata        \\
13 &2452034.4318& 23.82$\pm$0.08 & \nodata        \\ 14 &2452039.5863&
24.48$\pm$0.14 & 23.91$\pm$0.33 \\ 15 &2452044.4706& 24.21$\pm$0.11 &
23.90$\pm$0.35 \\ 16 &2452049.5547& 24.96$\pm$0.21 & $\geq$24.50    \\
17 &2452054.4373& 25.11$\pm$0.21 & $\geq$24.50    \\ 18 &2452059.5207&
25.18$\pm$0.24 & $\geq$24.50    \\ 19 &2452063.4651& 25.70$\pm$0.40 &
\nodata        \\ \enddata
\end{deluxetable}

\clearpage

\begin{deluxetable}{cccccc}
\tablecaption{Properties of Light Curves for Novae in
M49\label{tab:decline}} \tablewidth{0pt} \tablehead{ Nova  & $t_{\rm
max}$   & $V_{\rm max}$ & $M_{V,{\rm max}}$ & $t_2$  & $\nu_d$ \\ &
(Julian Date)   &     (mag)     &       (mag)       & (days) &
(days$^{-1}$) } \startdata 1 & $\le$2452008.4075   &
$\le$23.60$\pm$0.06 & $\le-7.53\pm0.12$ &  7.78$\pm$0.62 &
0.257$\pm$0.020 \\ 2 & $\le$2452008.4075   & $\le$22.64$\pm$0.04 &
$\le-8.49\pm0.11$ &  9.37$\pm$0.38 & 0.213$\pm$0.009 \\ 3 &
$\le$2452008.4075   & $\le$23.36$\pm$0.05 & $\le-7.77\pm0.11$ &
4.48$\pm$0.40 & 0.446$\pm$0.040 \\ 4 & 2452039.04$\pm$1.11 &
23.33$\pm$0.13      & $-7.80\pm0.16$    & 52.00$\pm$4.21 &
0.038$\pm$0.003 \\ 5 & 2452041.42$\pm$1.95 & 23.10$\pm$0.52      &
$-8.03\pm0.53$    &  7.53$\pm$1.14 & 0.266$\pm$0.011 \\ 6 &
2452023.46$\pm$0.50 & 23.33$\pm$0.28      & $-7.80\pm0.30$    &
19.30$\pm$0.86 & 0.104$\pm$0.005 \\ 7 & 2452027.33$\pm$0.50 &
22.72$\pm$0.36      & $-8.41\pm0.37$    & 11.68$\pm$0.83 &
0.171$\pm$0.012 \\ 8 & $\le$2452008.4075   & $\le$22.82$\pm$0.04 &
$\le-8.31\pm0.11$ & 17.71$\pm$0.81 & 0.113$\pm$0.005 \\ 9 &
2452031.45$\pm$0.50 & 23.74$\pm$0.22      & $-7.39\pm0.24$    &
37.16$\pm$2.86 & 0.054$\pm$0.004 \\ \enddata
\end{deluxetable}

\clearpage

\begin{deluxetable}{ccccccccccc}
\tablecaption{Absolute and Normalized Nova Rates\label{tab:novarate}}
\scriptsize \rotate \tablewidth{0pt} \tablehead{ Galaxy & T &
$\eta$\tablenotemark{a} & B & K$_{2MASS}$ & ($B-K$)$_{\rm
0}$\tablenotemark{b} & $A(B)_i$\tablenotemark{c} &
$A(B)_g$\tablenotemark{c} & ($m-M$)\tablenotemark{d} & $\nu_{\rm K,B}$
& $\nu_{\rm K,2MASS}$ \\ & & (year$^{-1}$) & (mag) & (mag) & (mag) &
(mag) & (mag) & (mag) & (year$^{-1}$~10$^{-10}$$L_{\rm K {\odot}})$ &
(year$^{-1}$~10$^{-10}$$L_{\rm K {\odot}}$)  } \startdata 
LMC &  9 & 2.5$\pm$0.5  & 0.91$\pm$0.05 &\nodata       & 2.74$\pm$0.10 & 0.07 &
0.32 & 18.50$\pm$0.13 & 5.98$\pm$1.26 & \nodata \\  
SMC &  9 & 0.3$\pm$0.2  & 2.70$\pm$0.10 &\nodata       & 2.71$\pm$0.10 & 0.24 &
0.13 & 18.99$\pm$0.05 & 2.50$\pm$1.68 & \nodata \\ 
M33 &  6 & 4.6$\pm$0.9  & 6.27$\pm$0.03 &4.11$\pm$0.04 & 2.87$\pm$0.10 & 0.33 &
0.18 & 24.64$\pm$0.09 & 4.27$\pm$0.89 & 12.6$\pm$2.7 \\ 
     &    &  $<$ 0.45      &               &              &               &      & &
& $<$ 0.41      & $<$ 1.22\\ 
M101 &  6 &    12$\pm$4 & 8.31$\pm$0.09 & 5.51$\pm$0.05& 3.24$\pm$0.11 & 0.05 & 0.04 & 29.34$\pm$0.10 &
1.01$\pm$0.35 & 1.63$\pm$0.57 \\ 
M51 &  4 & 18$\pm$7    & 8.96$\pm$0.06 & 5.05$\pm$0.03& 3.43$\pm$0.10 & 0.30 & 0.16 &
29.42$\pm$0.27 & 1.53$\pm$0.61 & 1.44$\pm$0.67 \\ 
M100 &  4 & 25$\pm$12.5 &10.05$\pm$0.08 & 6.59$\pm$0.04& 3.84$\pm$0.20 & 0.10 &
0.11 & 31.04$\pm$0.09 & 1.12$\pm$0.57 & 1.90$\pm$0.97 \\  
M31 &  3 & 29$\pm$4    & 4.36$\pm$0.02 & 0.98$\pm$0.02& 3.85$\pm$0.10 & 0.67 &
0.35 & 24.42$\pm$0.10 & 1.45$\pm$0.28 & 5.25$\pm$0.88 \\ 
M81 &  2 & 24$\pm$8    & 7.89$\pm$0.03 & 3.83$\pm$0.02& 3.99$\pm$0.10 & 0.34 &
0.35 & 27.80$\pm$0.08 & 1.64$\pm$0.56 & 2.73$\pm$0.94 \\ 
NGC5128 & -2 &    28$\pm$7    & 7.84$\pm$0.06 & 3.94$\pm$0.02& 3.38$\pm$0.11 & 0.00
& 0.50 & 28.12$\pm$0.14 & 2.84$\pm$0.76 & 2.67$\pm$0.75 \\ 
NGC1316 & -2 &   135$\pm$45   & 9.42$\pm$0.08 & 5.59$\pm$0.02& 4.15$\pm$0.20 &
0.00 & 0.09 & 31.66$\pm$0.17 & 1.61$\pm$0.55 & 2.33$\pm$0.86 \\ 
M87 & -4 &    91$\pm$34   & 9.59$\pm$0.04 & 5.81$\pm$0.02& 4.17$\pm$0.10 &
0.00 & 0.10 & 31.03$\pm$0.16 & 2.20$\pm$0.83 & 3.43$\pm$1.38 \\
VirgoEs & -4 &   160$\pm$57   & 9.46$\pm$0.10 & \nodata      & 4.26$\pm$0.11 & 0.00 & 0.09 & 31.17$\pm$0.09 & 2.79$\pm$1.03 & \nodata \\ 
M49 & -5 &   100$^{+35}_{-30}$  & 9.37$\pm$0.06 & 5.40$\pm$0.03& 4.30$\pm$0.10 & 0.00 & 0.09 & 31.06$\pm$0.10 & 1.71$\pm$0.61 & 2.52$\pm$0.91 \\
\enddata \tablenotetext{a}{Nova Rate References:  LMC: Capaccioli et
al. (1990).  SMC:  Graham (1979).  M33: Della Valle et al. (1994) and
Sharov (1993).  M101, M51: Shafter et al. (2000).  M100: Ferrarese et
al. (1996).  M31: Capaccioli et al. (1989).  M81: Moses \& Shafter
(1993).  NGC5128: Ciardullo et al. (1990).  NGC1316: Della Valle \&
Gilmozzi (2002).  M87: Shafter et al. (2000).  Virgo: Pritchet \& van
den Bergh (1987).  M49: this paper.  }
\tablenotetext{b}{Color References: LMC, SMC, M33, M101, M51, M31,
M81, NGC5128, M87, Virgo: Shafter et al. (2000).  M100: Aaronson
(1978).  NGC1316: Della Valle (2002).  M49: ($B-V$)$_0$ from NED and
($V-K$)$_0$ from Frogel et~al. (1978).  } \tablenotetext{c}{Errors on
the internal and Galactic extinction estimates are assumed to be 20\%
and 16\% respectively (the latter from Schlegel et al. 1998).  }
\tablenotetext{d}{Distance References:  LMC, SMC, M33, M101, M31, M81,
M100: Ferrarese et al. (2000).  M51, NGC5128, NGC1316, M87, M49,
Virgo: Tonry et al. (2001).  }
\end{deluxetable}

\end{document}